\documentclass[fleqn,twoside]{article}
\usepackage{amsmath,amssymb} 
\usepackage[headings]{espcrc2}

\readRCS
$Id: espcrc2.tex,v 1.2 2004/02/24 11:22:11 spepping Exp $
\newsavebox{\uuunit}
\sbox{\uuunit}
    {\setlength{\unitlength}{0.825em}
     \begin{picture}(0.6,0.7)
        \thinlines
        \put(0,0){\line(1,0){0.5}}
        \put(0.15,0){\line(0,1){0.7}}
        \put(0.35,0){\line(0,1){0.8}}
       \multiput(0.3,0.8)(-0.04,-0.02){10}{\rule{0.5pt}{0.5pt}}
     \end {picture}}
\newcommand {\unity}{\mathord{\!\usebox{\uuunit}}}
\title{BPS Black Holes}

\author{Bernard de Wit\address[ITPUU]{Institute for Theoretical
        Physics \& Spinoza Institute, \\ 
        Utrecht University, Utrecht, The Netherlands}%
} %
       
\runtitle{BPS Black Holes}
\runauthor{B. de Wit}
\begin{document}
\begin{abstract}
  The entropy of BPS black holes in four space-time dimensions is
  discussed from both macroscopic and microscopic points of view.
  \vspace{1pc}
\end{abstract}

\maketitle
\section{INTRODUCTION}
\label{sec:introduction}
Classical black holes are solutions of Einstein's equations of general
relativity that exhibit an event horizon. From inside this horizon,
nothing (and in particular, no light) can escape. The region inside
the horizon is therefore not in the backward lightcone of future
timelike infinity. However, since the discovery of Hawking radiation
\cite{Hawking}, it has become clear that many of the classical features
of black holes will be subject to change.

In these lecture we consider static, spherically symmetric black
holes in four space-time dimensions that carry electric and/or
magnetic charges with a flat space-time geometry at spatial infinity.
Such solutions exist in Einstein-Maxwell theory, the classical field
theory of gravity and electromagnetism. The most general static black
holes of this type correspond to the Reissner-Nordstrom solutions and
are characterized by a charge $Q$ and a mass $M$. In the presence of
magnetic charges, $Q$ is replaced by $\sqrt{q^2+p^2}$ in most
formulae, where $q$ and $p$ denote the electric and the magnetic
charge, respectively. Hence there is no need to distinguish between
the two types of charges. For zero charges one is dealing with
Schwarzschild black holes.

Two quantities associated with the black hole horizon are its area $A$
and the surface gravity $\kappa_{\rm s}$. The area is simply the area
of the two-sphere defined by the horizon. The surface gravity, which
is constant on the horizon, is related to the force (measured at
spatial infinity) that holds a unit test mass in place. On the other
hand, the mass $M$ and charge $Q$ of the black hole are not primarily
associated with the horizon and are defined in terms of appropriate
surface integrals at spatial infinity.

As is well known, there exists a striking correspondence between the
laws of thermodynamics and the laws of black hole mechanics
\cite{BCH}. Of particular importance is the first law, which, for
thermodynamics, states that the variation of the total energy is equal
to the temperature times the variation of the entropy, modulo work
terms, for instance proportional to a change of the volume. The
corresponding formula for black holes expresses how the variation of
the black hole mass is related to the variation of the horizon area,
up to work terms proportional to the variation of the angular
momentum. In addition there can also be a term proportional to a
variation of the charge, multiplied by the electric/magnetic potential
$\mu$ at the horizon. Specifically, the first law of thermodynamics,
$\delta E = T\,\delta S - p\,\delta V$, translates into
\begin{equation}
\label{first-law} 
\delta M = \frac{\kappa_{\rm s}}{2\pi} \, \frac {\delta A}{4}
+\mu\,\delta Q +\Omega\,\delta J\,. 
\end{equation} 
The reason for factorizing the first term on the right-hand side in
this particular form, is that $\kappa_{\rm s}/2\pi$ equals the Hawking
temperature \cite{Hawking}. This then leads to the identification of
the black hole entropy in terms of the horizon area,
\begin{equation}
\label{area-law}
{\cal S}_{\rm macro} = \tfrac14 A\,,
\end{equation}
a result that is known as the area law \cite{Bek}.
In these equations the various quantities have been defined in Planck
units, meaning that they have 
been made dimensionless by multiplication with an appropriate power of
Newton's constant (we will set $\hbar=c=1$). This constant appears in the
Einstein-Hilbert  
Lagrangian according to ${\cal L}_{\rm EH} = -(16\pi\,G_{\rm N})^{-1}\,
\sqrt{\vert g\vert} \,R$. With this normalization the quantities
appearing in the first law are independent of the scale of the
metric. 

Einstein-Maxwell theory can be naturally embedded into $N=2$
supergravity. This supergravity theory has possible extensions with
several abelian gauge fields and a related number of massless scalar
fields (often called `moduli' fields, for reasons that will become
clear later on).  At spatial infinity these moduli fields will tend to
a constant, and the black hole mass will depend on these constants,
thus introducing additional terms on the right-hand side of
(\ref{first-law}).

For Schwarzschild black holes the only relevant parameter is the mass
$M$ and we note the following relations, 
\begin{equation}
\label{schwarzschild}
 A= 16\pi\,M^2\,,\qquad \kappa_{\rm s}= \frac{1}{4\,M}\;,
\end{equation}
consistent with (\ref{first-law}). For the Reissner-Nordstrom black
hole, the situation is more subtle. Here one distinguishes three
different cases. For $M>Q$ one has non-extremal solutions, which
exhibit two horizons, an exterior event horizon and an interior
horizon. When $M=Q$ one is dealing with an extremal black hole, for
which the two horizons coalesce and the surface gravity vanishes. In
that case one has
\begin{equation}
\label{e-RN}
 A= 4\pi\,M^2\,,\qquad \kappa_{\rm s}=0\,,\qquad \mu = Q\,
 \sqrt{\frac{4\pi}A}\,.  
\end{equation}
It is straightforward to verify that this result is consistent with
(\ref{first-law}) for variations $\delta M=\delta Q$ and
$\kappa_{\mathrm{s}}=0$. Because the surface gravity vanishes, one
might expect the entropy to vanish as well, as is suggested by the
third law of thermodynamics. However, this is not the case because the
horizon area remains finite for zero surface gravity.  Finally,
solutions with $M<Q$ are not regarded as physically acceptable. Their
total energy is less than the electromagnetic energy alone and they no
longer have an event horizon but exhibit a naked singularity. Hence
extremal black holes saturate the bound $M\geq Q$ for physically
acceptable black hole solutions.

When embedding Einstein-Maxwell theory into a complete supergravity
theory, the above classification has an interpretation in terms of the
supersymmetry algebra. This algebra has a central extension
proportional to the black hole charge(s). Unitary representations of
the supersymmetry algebra must necessarily have masses that are larger
than or equal to the charge. When this bound is saturated, one is
dealing with so-called BPS supermultiplets. Such supermultiplets are
smaller than the generic massive $N=2$ supermultiplets and have a
different spin content. Because of this, BPS states are stable under
(adiabatic) changes of the coupling constants and the relation
between charge and mass remains preserved. This important feature of
BPS states will be relevant for what follows. In these lectures BPS
black hole solutions are defined by the fact that they have some
residual supersymmetry, so that they saturate a bound implied by the
supersymmetry algebra.

So far we did not refer to the explicit Schwarzschild or
Reissner-Nordstrom black hole solutions, which can be found in many
places in the literature. One feature that should be stressed,
concerns the near-horizon geometry. For extremal, static and
spherically symmetric black holes, this geometry is restricted to the
product of the sphere $S^2$ and an anti-de Sitter space
$\mathrm{AdS}_2$, corresponding to the line element,
\begin{equation}
  \label{eq:near-horizon}
  \mathrm{d}s^2 = - r^2 \mathrm{d}t^2 +
  \frac{\mathrm{d}r^2 + r^2(\mathrm{d}\theta^2 + \sin^2\theta
  \, \mathrm{d}\varphi^2)}{r^2} \,.
\end{equation}
In these coordinates the horizon is located at $r=0$, where the
timelike Killing vector $K = \partial_t$ turns lightlike. Such a
horizon is called a Killing horizon.

In these notes we discuss various aspects of the relation between
black hole solutions and corresponding microscopic descriptions.
Section \ref{sec:macro-micro} generally describes the macroscopic
(field theoretic) and microscopic (statistical) approach to black hole
entropy and indicates why they are related. Section \ref{sec:M-bh}
summarizes the calculation of the black hole entropy based on a
fivebrane wrapping on a Calabi-Yau four-cycle in a compactification of
M-Theory on the product space of a Calabi-Yau threefold and a circle.
In section \ref{sec:entropy-functions} the attractor equations are
discussed for extremal black holes on the basis of a variational
principle defined for a generic gravitational theory. Section
\ref{sec:n=2-supergravity} contains a brief review of $N=2$
supergravity and the superconformal multiplet calculus. This material
is used in the description of the BPS attractor equations in $N=2$
supergravity presented in section \ref{sec:bps-attractor-equations}.
In this case there also exists is a corresponding formulation in terms
of a BPS entropy function. Finally section
\ref{sec:partition-functions} discusses the relation between this
entropy function and the black hole partition function. 

We try to concentrate as much as possible on the conceptual basis of
the underlying ideas, but these notes cannot do justice to all aspects
relevant for the study of black holes in string theory, such as use of
the $\mathrm{AdS}_3/\mathrm{CFT}_2$ correspondence, black holes or
related objects in other than four space-time dimensions, the relation
with indices for BPS states, and the like. We refer to some recent
reviews where some of these topics have been discussed
\cite{Pioline:2006,Larsen:2006,Kraus:2006,Mohaupt:2007}.

\section{DUAL PERSPECTIVES}
\label{sec:macro-micro}
A central question in black hole thermodynamcis concerns the
statistical interpretation of the black hole entropy. String theory
has provided new insights here \cite{Strominger:1996sh}, which enable
the identification of the black hole entropy as the logarithm of the
degeneracy of states $d(Q)$ of charge $Q$ belonging to a certain
system of microstates. In string theory these microstates are provided
by the states of wrapped brane configurations of given momentum and
winding. When calculating the black hole solutions in the orresponding
effective field theory with the charges specified by the brane
configuration, one discovers that the black hole area is equal to the
logarithm of the brane state degeneracy, at least in the limit of
large charges. We will be reviewing some aspects of this remarkable
correspondence here.

The horizon area, which is expected to be proportional to the macroscopic
entropy according to the Bekenstein-Hawking area law, turns out to grow
quadratically with the charges $Q$. After converting to string units
the radius of a black hole is therefore proportional to 
\begin{equation}
  \label{eq:bh-radius}
  \frac{R_{\mathrm{horizon}}}{l_{\mathrm{string}}} \sim 
  g_{\mathrm{s}} \, Q\,, 
\end{equation}
where $l_{\mathrm{string}}$ and $g_{\mathrm{s}}$ are the string length
and coupling constant, respectively. Since we will be assuming that
the charges are large, the black holes are generically much larger
than the string scale. Consequently these black holes are called {\it
  large} and can be identified with the macroscopic black holes we
have been discussing earlier. However, there are also situations where
the leading contribution to the area is only linear in the charges
$Q$. In that case \eqref{eq:bh-radius} is replaced by
\begin{equation}
  \label{eq:bh-radius-small}
  \frac{R_{\mathrm{horizon}}}{l_{\mathrm{string}}} \sim 
  g_{\mathrm{s}} \, \sqrt{Q} \,, 
\end{equation}
Moreover, in that case the string coupling (inversely proportional to
the dilaton field) cannot be taken constant, but tends to zero for
large charges according to $g_{\mathrm{s}} \sim Q^{-1/2}$, so that the
radius of the black hole remains comparable to the string scale. These
black holes are called {\it small}. Their corresponding classical
supergravity solutions exhibit a vanishing horizon area and a dilaton
field that diverges at the horizon. To reliably compute the right-hand
side of \eqref{eq:bh-radius-small} therefore requires to include
appropriate terms in the effective action of higher order in
space-time derivatives. We return to this issue in subsection
\ref{sec:partial-legendre}.

To further understand the relation between a field-theoretic
description and a microscopic description it is relevant that strings
live in more that four space-time dimensions. In most situations the
extra dimensions are compactified on some internal manifold $X$ and
one is dealing with the standard Kaluza-Klein scenario leading to
effective field theories in four dimensions, describing low-mass modes
of the fields associated with appropriate eigenfunctions on the
internal manifold. Locally, the original space-time is a product
$M^4\times X$, where $M^4$ denotes the four-dimensional space-time
that we experience in daily life. In this situation there exists a
corresponding space $X$ at every point $x^\mu$ of $M^4$, whose size is
such that it will not be directly observable.  However, this space $X$
does not have to be the same at every point in $M^4$, and moving
through $M^4$ one may encounter various spaces $X$ that are not
necessarily equivalent. In principle they belong to some well-defined
class of fixed topology parametrized by certain moduli. These moduli
will appear as fields in the four-dimensional effective field theory.
For instance, suppose that the spaces $X$ are $n$-dimensional tori
$T^n$. The metric of $T^n$ will appear as a field in the
four-dimensional theory and is related to the torus moduli.  Hence,
when dealing with a solution of the four-dimensional theory that is
not constant in $M^4$, each patch in $M^4$ has a non-trivial image in
the space of moduli that parametrize the internal spaces $X$.

Let us now return to a black hole solution, viewed in this
higher-dimensional perspective. The fields, and in particular the
four-dimensional space-time metric, will vary nontrivially over $M^4$,
and so will the internal spaces $X$.  When moving to the center of the
black hole the gravitational fields will become strong and the
local product structure into $M^4\times X$ could break down.
Conventional Kaluza-Klein theory does not have much to say about what
happens, beyond the fact that the four-dimensional solution can be
lifted to the higher-dimensional one, at least in principle.

However, there is a feature of string theory that is absent in a
purely field-theoretic approach. In the effective field-theoretic
context only the local degrees of freedom of strings and branes are
captured. But extended objects may also carry global degrees of
freedom, as they can also wrap themselves around non-trivial cycles of
the internal space $X$. This wrapping tends to take place at a
particular position in $M^4$, so in the context of the four-dimensional
effective field theory this will reflect itself as a pointlike
object. The wrapped object is the string theory representation of the
black hole!

We are thus dealing with two complementary pictures of the black hole.
One is based on general relativity where a point mass generates a global
solution of space-time with strongly varying gravitational fields,
which we shall refer to as the {\it macroscopic} description.  The
other one, based on the internal space where an extended object is
entangled in one of its cycles, does not immediately involve
gravitational fields and can easily be described in flat space-time.
This description will be refered to as {\it microscopic}. To
understand how these two descriptions are related is far from easy,
but a connection must exist in view of the fact that gravitons are
closed string states which interact with the wrapped branes. These
interactions are governed by the string coupling constant $g_{\rm s}$
and we are thus confronted with an interpolation in that coupling
constant. In principle, such an interpolation is very difficult to
carry out, so that a realistic comparison between microscopic and
macroscopic results is usually impossible. However, reliable
predictions are possible for extremal black holes that are BPS! As we
indicated earlier, in that situation there are reasons to trust such
interpolations. Indeed, it has been shown that the predictions based
on these two alternative descriptions can be successfully compared and
new insights about black holes can be obtained.

But how do the wrapped strings and branes represent themselves in the
effective action description and what governs their interactions with
the low-mass fields? Here it is important to realize that the massless
four-dimensional fields are associated with harmonic forms on $X$.
Harmonic forms are in one-to-one correspondence with so-called
cohomology groups consisting of equivalence classes of forms that are
closed but not exact. The number of independent harmonic forms of a
given degree is given by the so-called Betti numbers, which are fixed
by the topology of the spaces $X$. When expanding fields in a
Kaluza-Klein scenario, the number of corresponding massless fields can
be deduced from an expansion in terms of tensors on $X$
corresponding to the various harmonic forms. The higher-dimensional
fields $\Phi(x,y)$ thus decompose into the massless fields $\phi^A(x)$
according to (schematically),
\begin{equation}
\label{KK-decomposition}
\Phi(x,y) = \phi^A(x)\;\omega_A(y)\,,
\end{equation}
where $\omega_A(y)$ denotes the independent harmonic forms on $X$. The above
expression, when substituted into the action of the higher-dimensional
theory, leads to interactions of the fields $\phi^A$ proportional to
the `coupling constants', 
\begin{equation}
\label{interaction}
C_{ABC\cdots} \propto \int_X \omega_A\wedge\omega_B\wedge\omega_C
\cdots\,. 
\end{equation}
These constants are known as intersection numbers, for reasons that
will become clear shortly. 

We already mentioned that the Betti numbers depend on the topology of
$X$. This is related to Poincar\'e duality, according to which
cohomology classes are related to homology classes. The latter consist
of submanifolds of $X$ without boundary that are themselves not a
boundary of some other submanifold of $X$. This is precisely relevant
for wrapped branes which indeed cover submanifolds of $X$, but are not
themselves the boundary of a submanifold because otherwise the brane
could collapse to a point. Without going into detail, this implies
that there exists a dual relationship between harmonic $p$-forms
$\omega$ and $(d_X-p)$-cycles, where $d_X$ denotes the dimension of
$X$. We can therefore choose a homology basis for the $(d_X-p)$-cycles
dual to the basis adopted for the $p$-forms.  Denoting this basis by
$\Omega_A$ the wrapping of an extended object can now be characterized
by specifying its corresponding cycle ${\cal P}$ in terms of the homology
basis,
\begin{equation}
\label{intersection}
{\cal P} = p^A\,\Omega_A\,.
\end{equation}
The integers $p^A$ count how many times the extended object is wrapped
around the corresponding cycle, so we are actually dealing with
integer-valued cohomology and homology. The wrapping numbers $p^A$
reflect themselves as magnetic charges in the effective action. The
electric charges are already an integer part of the effective action,
because they are associated with gauge transformations that usually
originate from the higher-dimensional theory. 

Owing to Poincar\'e duality it is thus very natural that the winding
numbers interact with the massless modes in the form of magnetic
charges, so that they can be incorporated in the effective action.
Before closing this section, we note that, by Poincar\'e duality, we
can express the number of intersections by
\begin{equation}
\label{intersections} 
P\cdot P\cdot P\cdots =C_{ABC\cdots}\,p^Ap^Bp^C\cdots\;.
\end{equation}
This is a topological characterization of the wrapping, which will
appear in later formulae.

\section{BLACK HOLES IN M/STRING THEORY -- AN EXAMPLE}
\label{sec:M-bh} 
As an example we now discuss the black hole entropy derived from
microscopic arguments in a special case. In a later section we will
consider the corresponding expression for the macroscopic entropy. We
start from M-theory, which is the strong coupling limit of type-IIA
string theory. Its massless states are described by eleven-dimensional
supergravity. The latter is invariant under 32 supersymmetries. Seven
of the eleven space-time dimensions are compactified on an internal
space which is the product of a Calabi-Yau threefold (a Ricci flat
three-dimensional complex manifold, henceforth denoted by $CY_3$)
times a circle $S^1$.  Such a space induces a partial breaking of
supersymmetry which leaves 8 supersymmetries unaffected. In the
context of the four-dimensional space-time $M^4$, these 8
supersymmetries are encoded into two independent Lorentz spinors and
for that reason this symmetry is referred to as $N=2$ supersymmetry.
Hence the effective four-dimensional field theory will be some version
of $N=2$ supergravity.

M-theory contains a five-brane and this is the microscopic object that
is responsible for the black holes that we consider: the five-brane
has wrapped itself on a 4-cycle ${\cal P}$ of the ${CY}_3$ space
\cite{Maldacena:1997de}.  Alternatively one may study this class of
black holes in type-IIA string theory, with a 4-brane wrapping the
4-cycle \cite{Vafa:1997gr}. The 4-cycle is subject to certain
requirements which will be mentioned in due course.

The massless modes captured by the effective field theory correspond
to harmonic forms on the $CY_3$ space; they do not depend on the $S^1$
coordinate. The 2-forms are of particular interest. In the effective
theory they give rise to vector gauge fields $A_\mu{}^A$, which
originate from the rank-three tensor gauge field in eleven dimensions.
In addition there is an extra vector field $A_\mu{}^0$ corresponding
to a 0-form which is related to the graviphoton associated with $S^1$.
This field will couple to the electric charge $q_0$ associated with
momentum modes on $S^1$ in the standard Kaluza-Klein fashion. The
2-forms are dual to 4-cyles and the wrapping of the five-brane is
encoded in terms of the wrapping numbers $p^A$, which appear in the
effective field theory as magnetic charges coupling to the gauge
fields $A_\mu{}^A$. Here we see Poincar\'e duality at work, as the
magnetic charges couple nicely to the corresponding gauge fields.  In
view of the fact that the product of three 2-forms defines a 6-form
that can be integrated over the $CY_3$ space, there exist non-trivial
triple intersection numbers $C_{ABC}$. These numbers will appear in
the three-point couplings of the effective field theory. There is a
subtle topological feature that we have not explained before, which is
typical for complex manifolds containing 4-cycles, namely the
existence of another quantity of topological interest known as the
second Chern class. The second Chern class is a 4-form whose integral
over a four-dimensional Euclidean space defines the instanton number.
The 4-form can be integrated over the 4-cycle ${\cal P}$ and yields
$c_{2A} \,p^A$, where the $c_{2A}$ are integers.

Let us now turn to the microscopic counting of degrees of freedom
\cite{Maldacena:1997de}.  These degrees of freedom are associated with
the massless excitations of the wrapped five-brane characterized by
the wrapping numbers $p^A$ on the 4-cycle. The 4-cycle ${\cal P}$ must
correspond to a holomorphically embedded complex submanifold in order
to preserve 4 supersymmetries.  The massless excitations of the
five-brane are then described by a $(1+1)$-dimensional superconformal
field theory (the reader may also consult \cite{MinMooTsi}).  Because
we have compactified the spatial dimension on $S^1$, we are dealing
with a closed string with left- and right-moving states. The 4
supersymmetries of the conformal field theory reside in one of these
two sectors, say the right-handed one.  Conformal theories in $1+1$
dimensions are characterized by a central charge, and in this case
there is a central charge for the right- and for the left-moving sector
separately. The two central charges are expressible in terms of the
wrapping numbers $p^A$ and depend on the intersection numbers and the
second Chern class, according to
\begin{eqnarray}
  \label{eq:c-LR}
  \begin{split}
    c_L &= C_{ABC}\, p^Ap^Bp^C + c_{2A}\,p^A \,, \\
    c_R &= C_{ABC}\, p^Ap^Bp^C + \tfrac12  c_{2A}\,p^A \,.
    \end{split}
\end{eqnarray}
Here we should stress that the above result is far from obvious and
holds only under the condition that the $p^A$ are large. In that case
every generic deformation of ${\cal P}$ will be smooth.  Under these
circumstances it is possible to relate the topological properties of
the 4-cycle to the topological data of the Calabi-Yau space.

We can now choose a state of given momentum $q_0$ which is
supersymmetric in the right-moving sector. From rather general
arguments it follows that such states exist. The corresponding states
in the left-moving sector have no bearing on the supersymmetry and
these states have a certain degeneracy depending on the value of
$q_0$. In this way we have a tower of degenerate BPS states invariant
under 4 supersymmetries, built on a supersymmetric state in the
right-moving sector. We can then use Cardy's formula, which states
that the degeneracy of states for fixed but large momentum (large as
compared to $c_L$) equals $\exp[2\pi\sqrt{\vert q_0\vert \, c_L/6}]$.
This leads to the following expression for the entropy,
\begin{equation}
  \label{eq:S-CY-micro}
  \begin{split}
   & {\cal S}_{\rm micro}(p,q) = \\
   &\quad 2\pi\sqrt{ \tfrac16 
  \vert\hat q_0\vert (C_{ABC} \,p^Ap^Bp^C + c_{2A}\,p^A)}\,,
  \end{split} 
\end{equation}
where $q_0$ has been shifted according to
\begin{equation}
  \label{eq:hat-q}
  \hat q_0 = q_0 + \tfrac1{2} C^{AB} q_Aq_B\,.
\end{equation}
Here $C^{AB}$ is the inverse of $C_{AB}= C_{ABC}p^C$.  This
modification is related to the fact that the electric charges
associated with the gauge fields $A_\mu{}^A$ will interact with the
M-theory two-brane \cite{Maldacena:1997de}. The existence of this
interaction can be inferred from the fact that the two-brane interacts
with the rank-three tensor field in eleven dimensions, from which the
vector gauge fields $A_\mu{}^A$ originate.

We stress that the above results apply in the limit of large charges.
The first term proportional to the triple intersection number is
obviously the leading contribution whereas the terms proportional to the
second Chern class are subleading. The importance of the subleading
terms will become more clear in later sections. Having obtained a
microscopic representation of a BPS black hole, it now remains to make
contact with it by deriving the corresponding black hole solution
directly in the $N=2$ supergravity theory. This is discussed in some
detail in section \ref{sec:bps-attractor-equations}. 

\section{ATTRACTOR EQUATIONS}
\label{sec:entropy-functions}
The microscopic expression for the black hole entropy depends only on
the charges, whereas a field-theoretic calculation can in principle
depend on other quantities, such as the values of the moduli fields at
the horizon. To establish any agreement between these two approaches,
the moduli (as well as any other relevant fields that enter the
calculation) must take fixed values at the horizon which may depend on
the charges. As it turns out this is indeed the case for extremal
black hole solutions, as was first demonstrated in the context of
$N=2$ supersymmetric black holes
\cite{Ferrara:1995ih,Strominger:1996kf,Ferrara:1996dd}. The values
taken by the fields at the horizon are independent of their asymptotic
values at spatial infinity. This fixed point behaviour is encoded in
so-called attractor equations.

In the presence of higher-derivative interactions it is very difficult
to explicitly construct black hole solutions and to exhibit the
attractor phenomenon.  However, by concentrating on the near-horizon
region one can usually determine the fixed-point values directly
without considering the interpolation between the horizon and spatial
infinity. Provided the symmetry of the near-horizon region is
restrictive enough, the attractor phenomenon can be described
conveniently in terms of a variational principle for a so-called
entropy function. The stationarity of this entropy function then
yields the attractor equations and its value at the attractor point
equals the macroscopic entropy. The purpose of the present section is
to explain this phenomenon.

For $N=2$ BPS black holes with higher-derivative interactions the
attractor equations follow from classifying possible solutions with
full supersymmetry \cite{LopesCardoso:2000qm}.  As it turns out
supersymmetry determines the near-horizon geometry (and thus the
horizon area), the values of the moduli fields in terms of the charges
and the value of the entropy as defined by the Noether charge
definition of Wald \cite{Wald:1993nt}. This result can also be
described in terms of a variational principle
\cite{Behrndt:1996jn,LopesCardoso:2006bg}, as we shall explain in
section \ref{sec:bps-attractor-equations}. 

Let us now turn to the more generic case of extremal, static,
spherically symmetric black holes that are not necessarily BPS,
following the approach of \cite{Sen:2005wa} which is based on an
action principle.  Similar approaches can be found elsewhere in the
literature, for instance, in
\cite{Ferrara:1997tw,Gibbons:1997cc,Goldstein:2005hq}. When dealing
with spherically symmetric solutions, one integrates out the spherical
degrees of freedom and obtains a reduced action for a $1+1$
dimensional field theory which fully describes the black hole
solutions. Here we consider a general system of abelian vector gauge
fields, scalar and other matter fields coupled to gravity. The
geometry is thus restricted to the product of the sphere $S^2$ and a
$1+1$ dimensional space-time, and the dependence of the fields on the
$S^2$ coordinates $\theta$ and $\varphi$ is fixed by symmetry
arguments. For the moment we will not make any assumption regarding
the dependence on the remaining two cooordinates $r$ and $t$.
Consequently we write the general field configuration consistent with
the various isometries as
\begin{equation}
  \label{eq:general-fields}
  \begin{split}
    \mathrm{d}s^2{}_{(4)} &= g_{\mu\nu}
  \mathrm{d}x^\mu\mathrm{d}x^\nu \\
  &=
  \mathrm{d}s^2{}_{(2)} 
  + v_2 \Big(\mathrm{d} \theta^2 +\sin^2\theta \,\mathrm{d}\varphi^2\Big)\,,
  \\ 
  F_{rt}{}^I &= e^I\,,\qquad F_{\theta\varphi}{}^I = \frac{p^I}{4\pi}\,
  \sin\theta \,.
  \end{split}
\end{equation}
The $F_{\mu\nu}{}^I$ denote the field strengths associated with a
number of abelian gauge fields. The $\theta$-dependence of
$F_{\theta\varphi}{}^I$ is fixed by rotational invariance and the
$p^I$ denote the magnetic charges. The latter are constant by virtue
of the Bianchi identity, but all other fields are still functions of
$r$ and $t$. As we shall see in a moment the fields $e^I$ are dual to
the electric charges. The radius of $S^2$ is defined by the field
$v_2$. The line element of the $1+1$ dimensional space-time will be
expressed in terms of the two-dimensional metric $\bar g_{ij}$, whose
determinant will be related to a field $v_1$ according to,
\begin{equation}
  \label{eq:v-1}
  v_1 = \sqrt {\vert\bar g\vert} \,.
\end{equation}
Eventually $\bar g_{ij}$ will be taken proportional to an $AdS_2$ metric,
\begin{equation}
  \label{eq:bar-g}
  \mathrm{d}s^2{}_{(2)} = \bar g_{ij} \,\mathrm{d}x^i\mathrm{d}x^j =  
  v_1\Big(-r^2\,\mathrm{d}t^2 + \frac{\mathrm{d} r^2}{r^2}\Big) \,.
\end{equation} 
In addition to the fields $e^I$, $v_1$ and $v_2$, there may be a number
of other fields which for the moment we denote collectively by
$u_\alpha$.

As is well known theories based on abelian vector fields are subject
to electric/magnetic duality, because their equations of motion
expressed in terms of the dual field
strengths,\footnote{
  We assume that the Lagrangian is a function of the abelian field
  strengths and does not depend explicitly on the gauge fields. }
\begin{equation}
  \label{eq:dual-F}
  G_{\mu\nu I} =  \sqrt{\vert g\vert}\,
  \varepsilon_{\mu\nu\rho\sigma}\, 
  \frac{\partial \mathcal{L}}{\partial F_{\rho\sigma}{}^I} \,,
\end{equation}
take the same form as the Bianchi identities for the field strengths
$F_{\mu\nu}{}^I$. Adopting the conventions where
$x^\mu=(t,r,\theta,\varphi)$ and $\varepsilon_{tr\theta\varphi}= 1$,
and the signature of the space-time metric equals $(-,+,+,+)$, which
is consistent with \eqref{eq:bar-g}, it follows that, in the background
\eqref{eq:general-fields},
\begin{eqnarray}
  \label{eq:G}
  \begin{split}
  G_{\theta\varphi\,I}&= 
  - v_1 v_2
  \,\sin\theta\, \frac{\partial\mathcal{L}} {\partial e^I}  \,,
  \\ 
  G_{rt\,I}&= 
   - 4\pi\,v_1v_2\,
  \frac{\partial\mathcal{L}} {\partial p^I}  \,.
  \end{split}
\end{eqnarray}
These two tensors can be written as $q_I\,\sin\theta/(4\pi)$ and
$f_I$. The quantities $q_I$ and $f_I$ are conjugate to $p^I$ and $e^I$,
respectively, and can be written as
\begin{eqnarray}
  \label{eq:def-q-f}
  \begin{split}
  q_I(e,p,v,u) &= - 4\pi\, v_1v_2 \,
  \frac{\partial\mathcal{L}} {\partial e^I}  \,, \\
  f_I(e,p,v,u) &= -  4\pi\,v_1v_2 \,
  \frac{\partial\mathcal{L}} 
  {\partial p^I}  \,. 
  \end{split}
\end{eqnarray}
They depend on the constants $p^I$ and on the fields $e^I$, $v_{1,2}$
and $u_\alpha$, and possibly their $t$ and $r$ derivatives, but no
longer on the $S^2$ coordinates $\theta$ and $\varphi$. Upon imposing
the field equations it follows that the $q_I$ are constant and
correspond to the electric charges. Our aim is to obtain a description
in terms of the charges $p^I$ and $q_I$, rather than in terms of the
$p^I$ and $e^I$.

Electric/magnetic duality transformation are induced by rotating the
tensors $F_{\mu\nu}{}^I$ and $G_{\mu\nu\,I}$ by a constant
transformation, so that the new linear combinations are all subject to
Bianchi identities. Half of them are then selected as the new field
strengths defined in terms of new gauge fields, while the Bianchi
identities on the remaining linear combinations are regarded as field
equations belonging to a new Lagrangian defined in terms of the new
field strengths. In order that this dualization can be effected the
rotation of the tensors must belong to $\mathrm{Sp}(2n+2;\mathbb{R})$,
where $n+1$ denotes the number of independent gauge fields
\cite{Gaillard:1981rj}. Naturally the duality leads to new quantities
$(\tilde p^I,\tilde q_I)$ and $(\tilde e^I,\tilde f_I)$, related to
the original ones by the same $\mathrm{Sp}(2n+2;\mathbb{R})$ rotation.
Since the charges are not continuous but will take values in an
integer-valued lattice, this group should eventually be restricted to
an appropriate arithmetic subgroup.
 
Subsequently we define the reduced Lagrangian by the integral of the
full Lagrangian over $S^2$, 
\begin{equation}
  \label{eq:reduced-action}
  \mathcal{F}(e,p,v,u) = \int \mathrm{d}\theta\,\mathrm{d}\varphi\;
  \sqrt{\vert g\vert} \, \mathcal{L} \,, 
\end{equation}
We note that the definition of the conjugate quantities $q_I$ and
$f_I$ takes the form, 
\begin{equation}
  \label{eq:q-F-f-F}
  q_I = - \frac{\partial\mathcal{F}}{\partial e^I}\;, \qquad
  f_I =  - \frac{\partial\mathcal{F}}{\partial p^I}\;.
\end{equation}

It is known that a Lagrangian does not transform as a function under
electric/magnetic dualities (see, e.g. \cite{deWit:2001pz}), but one
can generally show that the following combination does \cite{CdWM},
\begin{equation}
  \label{eq:entropy-function}
  \mathcal{E}(q,p,v,u) =- \mathcal{F}(e,p,v,u) -e^I q_I \,.
\end{equation}
More precisely, this quantity transforms under electric/magnetic
duality according to $\tilde {\mathcal{E}}(\tilde q,\tilde p, v,u) =
{\mathcal{E}}(q,p, v,u)$.  In view of the first equation
(\ref{eq:q-F-f-F}), the definition \eqref{eq:entropy-function} takes
the form of a Legendre transform.  Furthermore the field equations
imply that the $q_I$ are constant and that the action, $\int\,
\mathrm{d}t\mathrm{d}r\, \mathcal{E}$, is stationary under variations
of the fields $v$ and $u$, while keeping the $p^I$ and $q_I$ fixed.
This is to be expected as $\mathcal{E}$ is in fact minus the
Hamiltonian density associated with the reduced Lagrangian density
\eqref{eq:reduced-action}, at least as far as the vector fields are
concerned.

In the near-horizon background (\ref{eq:bar-g}), assuming fields that
are invariant under the $AdS_2$ isometries, the generally covariant
derivatives of the fields vanish and the function $\mathcal{E}$
depends only on the fields which no longer depend on the coordinates.
The equations of motion then imply that the values of the fields
$v_{1,2}$ and $u_{\alpha}$ are determined by demanding $\mathcal{E}$
to be stationary under variations of $v$ and $u$,
\begin{eqnarray}
  \label{eq:field-eqs}
  \frac{\partial \mathcal{E}}{\partial v} =  \frac{\partial
  \mathcal{E}}{\partial u} = 0\,,\qquad q_I = \mathrm{constant}\,.
\end{eqnarray}
The function $2\pi\,\mathcal{E}(q, p, v,u)$ then coincides with the entropy
function proposed by Sen \cite{Sen:2005wa}. The first two equations of
\eqref{eq:field-eqs} are interpreted as the attractor equations.
At the attractor point one may prove 
\begin{equation}
  \label{eq:entropy-F}
  \mathcal{E}\Big\vert_{\mathrm{attractor}} = - \int
  \mathrm{d}\theta\,\mathrm{d}\varphi \;\sqrt{\vert 
  g\vert} \, R_{rtrt}\,\frac{\partial \mathcal{L}}{\partial R_{rtrt}}\,, 
\end{equation}
where the right-hand side is evaluated in the near-horizon geometry.
This leads to the expression
\begin{equation}
  \label{eq:wald}
  2\pi\,\mathcal{E}\Big\vert_{\mathrm{attractor}} = 2\pi
  \int_{\Sigma_{\mathrm{hor}}} \, \frac{\partial \mathcal{L}}{\partial
  R_{\mu\nu\rho\sigma}} \,\varepsilon_{\mu\nu}\varepsilon_{\rho\sigma}
  \,,  
\end{equation}
where $\Sigma_{\mathrm{hor}}$ denotes a spacelike cross section of the
Killing horizon, and $\varepsilon_{\mu\nu}$ the normal bivector which
acts in the space normal to $\Sigma_{\mathrm{hor}}$.  This is
precisely the expression for the Wald entropy \cite{Wald:1993nt}
(applied to this particular case). For the Einstein-Hilbert action,
\eqref{eq:wald} equals a quarter of the horizon area in Planck units.
For more general Lagrangians \eqref{eq:wald} may lead to deviations
from the area law, as we will see in due course. Note that the entropy
function does not necessarily depend on all fields at the horizon. The
values of some of the fields will then be left unconstrained, but
those will not appear in the expression for the Wald entropy either.

The above derivation of the entropy function applies to any gauge and
general coordinate invariant Lagrangian, including Lagrangians with
higher-derivative interactions. In the absence of higher-derivative
terms, the reduced Lagrangian $\mathcal{F}$ is at most quadratic in
$e^I$ and $p^I$ and the Legendre transform \eqref{eq:entropy-function}
can easily be carried out. The results coincide with corresponding
terms in the so-called black hole potential discussed in e.g.
\cite{Ferrara:1997tw,Gibbons:1997cc,Goldstein:2005hq}.

\section{N=2 SUPERGRAVITY}
\label{sec:n=2-supergravity}
In the previous section the symmetry of the near-horizon geometry
played a crucial role. For BPS black holes the supersymmetry
enhancement at the horizon is the crucial input that constrains both
some of the fields at the horizon as well as the near-horizon geometry
itself.  The black hole solutions that we will be considering have a
residual $N=1$ supersymmetry (so that they are BPS) and are solitonic
interpolations between $N=2$ configurations at the horizon and at
spatial infinity.  Obviously to describe BPS black holes one needs to
consider extended supergravity theories. In view of the application
described in section \ref{sec:M-bh}, $N=2$ supergravity is relevant.
Moreover, $N=2$ supergravity has off-shell formulations (meaning that
supersymmetry transformations realize the supersymmetry algebra
without the need for imposing field equations associated with a
specific Lagrangian) and this facilitates the calculations in an
essential way.  This aspect is especially important because we will be
considering supersymmetric Lagrangians with interactions containing
more than two derivatives.

In the following subsections we present a brief introduction to $N=2$
supergravity. Supermultiplets are introduced in subsection
\ref{sec:supermultiplets} and supersymmetric actions in subsection
\ref{sec:supersymmetric-actions}. Finally, in subsection
\ref{sec:compensator-multiplets}, we elucidate the use of compensating
fields and corresponding supermultiplets to familiarize the reader
with the principles underlying the superconformal multiplet calculus.
Further details can be found in the literature \cite{DWVHVPL,BDRDW}.

\subsection{Supermultiplets}
\label{sec:supermultiplets}
In this subsection we briefly introduce the supermultiplets that play
a role in the following. Of particular interest are the vector and the
Weyl supermultiplet. Other multiplets are the tensor supermultiplets
and the hypermultiplets, but those will not be discussed as they only
play an ancillary role.

The covariant fields and field strengths of the various gauge fields
belonging to the vector or to the Weyl supermultiplet comprise a
chiral multiplet. Such multiplets are described in superspace by chiral
superfields.At this point it is convenient to systematize our
discussion by using superspace notions, although we do not intend to
make an essential use of superfields. Scalar chiral fields in $N=2$
superspace have $16+16$ bosonic $+$ fermionic field components, but
there exists a constraint which reduces the superfield to only $8+8$
field components. This constraint expresses higher-$\theta$ components
of the superfield in terms of lower-$\theta$ components or space-time
derivatives thereof. The vector supermultiplet and the Weyl
supermultiplet are both related to {\it reduced} chiral multiplets.
Note, however, that products of reduced chiral superfields constitute
general chiral fields.

The vector supermultiplet is described by a scalar reduced chiral
superfields, whose lowest-$\theta$ component is a complex field which
we denote by $X$. Then there is a doublet of chiral fermions
$\Omega_i$, where $i$ is an $\mathrm{SU}(2)$ R-symmetry doublet index.
The position of the index $i$ indicates the chirality of the spinor
field: $\Omega_i$ carries positive, and $\Omega^i$ negative chirality.
The fields $\bar X$ and $\Omega^i$ appear as lowest-$\theta$
components in the anti-chiral superfield that one obtains by complex
conjugation of the chiral superfield. We recall that the so-called
R-symmetry group is defined as the maximal group that rotates the
supercharges in a way that commutes with Lorentz symmetry and is
compatible with the supersymmetry algebra. For $N=2$ supersymmetry
this group equals $\mathrm{SU}(2)\times \mathrm{U}(1)$, which acts
chirally on the spinors. In spite of its name, R-symmetry does not
necessarily consitute an invariance of supersymmetric Lagrangians.
Finally, at the $\theta^2$-level, we encounter the field strength
$F_{\mu\nu}$ of the gauge field and an auxiliary field which we write
as a symmetric real tensor, $Y_{ij} = Y_{ji}= \varepsilon_{ik}
\varepsilon_{jl} Y^{kl}$.  Here we note that complex conjugation will
often be indicated by rasing and/or lowering of $\mathrm{SU}(2)$
indices. One can easily verify that in this way we have precisely
$8+8$ independent field components (which we will refer to as
off-shell degrees of freedom, as we did not impose any field
equations). Note the difference with on-shell degrees of freedom. The
conventional Lagrangian for the vector supermultiplet describes $4+4$
physical massless bosonic and fermionic states: 2 scalars associated
with $X$ and $\bar X$, the 2 helicities associated with the vector
gauge field, and 2 Majorana fermions, each carrying 2 helicities.

The Weyl supermultiplet has a rather similar decomposition, but in
this case the reduced chiral superfield is not a scalar but an
anti-selfdual Lorentz tensor. For extended supersymmetry the Weyl
superfield is also assigned to the antisymmetric representation of the
R-symmetry group, so that its lowest-order $\theta$-component is
denoted by $T_{ab}{}^{ij}$. Its complex conjugate belongs to the
corresponding anti-chiral superfield and its corresponding tensor is
selfdual and denoted by $T_{abij}$. Here the indices $a,b,\ldots$
denote the components of space-time tangent space tensors. In view of
its tensorial character the Weyl supermultiplet comprises $24+24$
off-shell degrees of freedom. The covariant components of the Weyl
multiplet are as follows. Linear in $\theta$ one has the fermions
decomposing into the field strength of the gravitini and a 
doublet spinor $\chi^i$. The gravitino field strengths comprise $16$
degrees of freedom so that together with the spinors $\chi^i$ we
count 24 off-shell degrees of freedom. At the $\theta^2$-level we have
the Weyl tensor, the field strengths belonging to the gauge fields
associated with R-symmetry, and a real scalar field denoted by $D$,
comprising 5, $4\times 3$ and 1 off-shell degrees of freedom,
respectively. Together with the 6 degrees of freedom belonging to
$T_{ab}{}^{ij}$, we count 24 bosonic degrees of freedom.

The Weyl multiplet contains the fields of $N=2$ conformal supergravity
and an invariant action can be written down that is quadratic in its
components. Although $T_{ab}{}^{ij}$ is not subject to a Bianchi
identity, it is often called the ``graviphoton field strength''. The
reason for this misnomer is that the gravitini transform into
$T_{ab}{}^{ij}$ and in locally supersymmetric Lagrangians of vector
multiplets that are at most quadratic in space-time derivatives,
$T_{ab}{}^{ij}$ acts as an auxiliary field and couples to a
field-dependent linear combination of the vector multiplet field
strengths. For this class of Lagrangians, all the fields of the Weyl
multiplet with the exception of the graviton and the gravitini fields,
act as auxiliary fields. 

In this subsection we mainly describe linearized results. Ultimately
we are interested in constructing a theory of local supersymmetry.
This means that the vector multiplets must first be formulated in a
supergravity background. This leads to additional terms in the
supersymmetry transformation rules and in the superfield components
which depend on the supergravity background. Some of these terms
correspond to replacing ordinary space-time derivatives by covariant
ones. However, we consider only vector multiplets and the Weyl
multiplet here and the latter describes the supermultiplet of {\it
  conformal} supergravity. Consistency therefore requires that we
formulate the vector supermultiplet in a superconformal background,
and, indeed, the vector supermultiplet is a representation of the full
superconformal algebra. Therefore all the superconformal symmetries
can be realized as local symmetries. Naturally the Weyl multiplet
itself will also acquire non-linear corrections but those do not
involve the vector multiplet fields. The reader may wonder why we are
interested in conformal supergravity here, but this will be clarified
shortly. Other than that, the situation is conceptually the same as
when considering multiplets of matter fields coupled to a nonabelian
gauge theory, or to gravity.

\subsection{Supersymmetric actions}
\label{sec:supersymmetric-actions}
In view of the fact that both the vector and Weyl supermultiplets are
described by chiral superfields, even beyond their linearized form, it
is clear how to construct supersymmetric actions. Namely one takes
some function of the vector superfield (actually we will need several
of these superfields, which we label by indices $I=0,1,2,\ldots,n$)
and the square of the Weyl superfield. Taking the square implies no
loss of generality because we are interested in Lorentz invariant
couplings. When expanding the superfields in $\theta$ components, one
generates multiple derivatives of this function which depend on the
lowest-$\theta$ components, $X$ and 
\begin{equation}
  \label{eq:A=T2}
  A= (T_{ab}{}^{ij}\varepsilon_{ij})^2\,.
\end{equation}
Because the function is holomorphic, i.e., it depends on $X$ and
$A$, but not on their complex conjugates, we take the imaginary part
of the resulting expression. However, in order that the action is
superconformally invariant, the function $F(X,A)$ must be holomorphic
and homogeneous,
\begin{equation}
  \label{eq:homo-F}
  F(\lambda X, \lambda^2 A) = \lambda^2 F(X, A)\,.
\end{equation}
We refrain from giving full results. In principle they are derived
straightforwardly, but the formulae are often lengthy and require
extra definitions. Therefore we discuss only a few characteristic
terms.

First of all, let us consider the scalar kinetic terms. They are
accompanied by a coupling to the Ricci scalar and the scalar field
$D$ of the Weyl multiplet in the following way,
\begin{equation}
  \label{eq:L_X}
  \begin{split}
    \mathcal{L} &\propto \mathrm{i}(\partial_\mu F_I \,\partial^\mu \bar
  X^I  - \partial_\mu \bar F_I \,\partial^\mu X^I) \\
  &\quad
  - \mathrm{i}(\tfrac16 R-D)(F_I\,\bar X^I-\bar F_I\,X^I) \,, 
  \end{split}
\end{equation}
where $F_I$ denotes the derivative of $F$ with respect to $X^I$.
Observe that when the function $F$ depends on $T_{ab}{}^{ij}$ this
will generate interactions between the kinetic term for the vector
multiplet scalars and the tensor field of the Weyl multiplet. Of
course, this pattern continues for other terms.

A second term concerns the kinetic term of the vector fields, which is
proportional to the second derivative of the function $F$, 
\begin{equation}
  \label{eq:L_FF}
  \begin{split}
    \mathcal{L}&\propto \tfrac14\mathrm{i}\,F_{IJ}\, (F^-_{ab}{}^I
  -\tfrac14 \bar X^I T_{ab}{}^{ij}\varepsilon_{ij} ) \,  \\
  &\quad
  \times 
  (F^{-ab J} -\tfrac14 \bar X^J T^{abkl}\varepsilon_{kl} ) 
  + \mathrm{h.c.}\,. 
  \end{split}
\end{equation}

A third term involves the square of the Weyl tensor, contained in
the tensor $\mathcal{R}(M)$, 
\begin{equation}
  \label{eq:L_RR}
  \begin{split}
  \mathcal{L}&\propto 16 \,\mathrm{i}\,F_{A}[2\, \mathcal{R}(M)^{cd}{}_{ab}
  \,\mathcal{R}(M)^-_{cd}{}^{ab} \\
  &\quad
  -16\, T^{abij}\,D_aD^c T_{cbij} ]   + \mathrm{h.c.} \,. 
  \end{split}
\end{equation}
Here $D_a$ denotes a superconformally covariant derivative (which also
contains terms proportional to the Ricci tensor). We refrain from
giving further details at this point and refer to the literature.

\subsection{Compensator multiplets}
\label{sec:compensator-multiplets}
The theories discussed so far are invariant under the local symmetries
of the superconformal algebra. This high degree of symmetry seems
unnecessary for, or even an obstacle to, practical applications. The
purpose of this section is to explain that this is not the case.

Let us start with a simple example, namely massive $\mathrm{SU}(N)$ 
Yang-Mills theory, with Lagrangian,
\begin{equation}
  \label{eq:massiveYM}
  \begin{split}
    \mathcal{L} &= \tfrac14 \mathrm{Tr}[ F_{\mu\nu}(W)\,F^{\mu\nu}(W)]
  \\ 
  &\quad 
  +\tfrac12 M^2 \mathrm{Tr}[ W_{\mu}\,W^{\mu}] \,,
  \end{split}
\end{equation}
where we use a Lie-algebra valued notation and the definition
$F_{\mu\nu} =2\, \partial_{[\mu} W_{\nu]} - [W_\mu,W_\nu]$. Introduce
now a matrix-valued field $\Phi\in \mathrm{SU}(N)$ transforming under
$\mathrm{SU}(N)$ gauge transformations from the left and substitute
$W_\mu = \Phi^{-1} D_\mu\Phi$ into the Lagrangian, where the covariant
derivative reads $D_\mu\Phi =(\partial_\mu-A_\mu)\Phi$. The first term
is not affected by this transformation which takes the form of a
field-dependent gauge transformation. But the mass term changes, and
we find the following Lagrangian,
\begin{equation}
  \label{eq:Stueck-YM}
  \begin{split}
  \mathcal{L}&= \tfrac14 \mathrm{Tr}[
  F_{\mu\nu}(A)\,F^{\mu\nu}(A)]  \\
  &\quad
  - \tfrac12 M^2\, \mathrm{Tr}[ D_\mu \Phi D^\mu \Phi^{-1}] \,,
  \end{split}
\end{equation}
which is manifestly gauge invariant. Clearly, this is a massless gauge
theory coupled to $N^2-1$ scalars. However, this formulation is gauge
equivalent to \eqref{eq:massiveYM}, as one readily verifies by imposing
the gauge condition $\Phi=\unity$.

What do we achieve by rewriting \eqref{eq:massiveYM} into the form
\eqref{eq:Stueck-YM}? Both Lagrangians describe the same number of
physical states and are based on the same number of off-shell degrees
of freedom. In \eqref{eq:massiveYM} the degrees of freedom are contained
in a single field, $W_\mu$, carrying 4 components per generator. In
\eqref{eq:Stueck-YM}, however, the degrees of freedom are distributed
over two fields in a local and Lorentz invariant way, namely 3
components per generator in $A_\mu$ (we subtracted the gauge degrees
of freedom) and 1 component per generator in $\Phi$. Hence the second
formulation can be regarded as reducible, and this reducibility has
been achieved by introducing extra gauge invariance.

A similar situation exists for gravity, as is shown by the Lagrangian,
\begin{equation}
  \label{eq:scale-inv-R}
  \sqrt{\vert g\vert} \mathcal{L}\propto \sqrt{\vert g\vert} \left[
  g^{\mu\nu} \partial_\mu\phi\, \partial_\nu\phi - \tfrac16 R\, \phi^2
  \right] \,, 
\end{equation}
which is invariant under local scale transformation with parameter
$\Lambda(x)$: $\delta\phi= \Lambda \,\phi$, $\delta g_{\mu\nu} = -2
\Lambda\, g_{\mu\nu}$. This Lagrangian is gauge equivalent to the
Einstein-Hilbert Lagrangian. To see this one either rewrites it in
terms of a scale invariant metric $\phi^2 g_{\mu\nu}$, or one simply
imposes the gauge condition and sets $\phi$ equal to a constant (which
will then be related to Newton's constant). Again the Einstein-Hilbert
Lagrangian and \eqref{eq:scale-inv-R} contain the same number of
off-shell degrees of freedom, but the latter field configuration is
reducible: $g_{\mu\nu}$ describes only 5 degrees of freedom (in view
of the local scale invariance) and the sixth one can be assigned to
the scalar field $\phi$.

Fields such as $\Phi$ and $\phi$ are called compensating fields,
because they can be used to convert any quantity that transforms under
the gauge symmetry into a gauge invariant one. Often the gauge
equivalent formulation, based on the introduction of compensator
fields and gauge symmetries at the same time, is exploited for reasons
of renormalizability as one can use the gauge freedom to choose a
different gauge that leads to better short-distance behaviour. This is
not the issue here but the crucial point is that the compensating
degrees of freedom must be contained in full supermultiplets. By
keeping the gauge invariance manifest one realizes a higher degree of
symmetry which facilitates the construction of Lagrangians and
clarifies the geometrical features of the resulting supergravity
theories. In this way, pure $N=2$ supergravity is, for instance,
constructed from two compensating supermultiplets, one of which is a
vector multiplet and the other one can be a tensor multiplet or a
hypermultipet. Of these two supermultiplet only the vector field will
describe physical degrees of freedom (namely, those corresponding to
the graviphoton). The other components play the role of either
compensating fields (associated with local scale, R-symmetry and
special conformal supersymmetry transformations), or are constrained by
the field equations, either by Lagrange multipliers, or because they
are auxiliary.

\section{BPS ATTRACTORS}
\label{sec:bps-attractor-equations}
As we already discussed in section \ref{sec:entropy-functions} the BPS
attractor equations follow from the requirement of full $N=2$
supersymmetry at the horizon. In the context of an off-shell
representation of superymmetry, the corresponding equations can be
derived in a way that is rather independent of the action. As
explained in the previous section, $N=2$ vector multiplets contain
complex physical scalar fields which we denoted by $X^I$. In the
context of the superconformal framework these fields are defined
projectively, in view of the invariance under local scale
and $\mathrm{U}(1)$ transformations. The action for vector
multiplets with additional interactions involving the square of the
Weyl tensor is encoded in a holomorphic function $F(X,A)$, which is
homogenous of second degree (c.f.  \eqref{eq:homo-F}). Here $A$ is
quadratic in the anti-selfdual field $T_{ab}{}^{ij}$, as shown in
\eqref{eq:A=T2}. The normalizations of the Lagrangian and of the
charges adopted below differ from the normalizations used in section
\ref{sec:entropy-functions}.

Another issue that we should explain concerns electric/magnetic
duality, which in principle pertains to the gauge fields.
Straightforward application of this duality to an $N=2$ supersymmetric
Lagrangian with vector multiplets, leads to a new Lagrangian that no
longer takes the canonical form in terms of a function $F(X,A)$. In
order to bring it into that form one must simultaneously apply a field
redefinition to the scalar and spinor fields. On the scalar fields,
this redefinition follows from the observation that $(X^I,F_I(X,A))$
transforms as a sympletic vector analogous to the tensors
$(F_{\mu\nu}{}^I, G_{\mu\nu I})$ discussed previously. Hence the
$(2n+2)$-dimensional vector $(X^I,F_I(X,A))$ is rotated into a new
vector $(\tilde X^I,\tilde F_I)$. When this rotation belongs to
$\mathrm{Sp}(2n+2;\mathbb{R})$ then $\tilde F_I$ can be written as the
derivative of a new function $\tilde F(\tilde X,A)$ with respect to
$\tilde X^I$. The new function then encodes the dual action. The need
for this additional field redefinition follows from the fact that the
gauge fields, and thus their field strengths $F_{\mu\nu}{}^I$, and the
fields $X^I$ have a well-defined relation encoded in the supersymmetry
transformation rules. Therefore, the transformations of
$(X^I,F_I(X,A))$ must be taken into account when considering
electric/magnetic duality.  We refer to
\cite{deWit:2001pz,deWit:1996ix} for further details and a convenient
list of formulae.
 
To determine the BPS attractor equations one classifies all possible
$N=2$ supersymmetric solutions. This is done by studying the
supersymmetry variations of the fermions in an arbitrary bosonic
background.  Requiring that these variations vanish then imposes
strong restrictions on this background. The analysis was performed in
\cite{LopesCardoso:2000qm} for $N=2$ supergravity with an arbitrary
number of vector multiplets and hypermultiplets, including
higher-order derivative couplings proportional to the square of the
Weyl tensor. Also the interpolating BPS solutions were studied in
considerable detail. It was found that $N=2$ supersymmetric solutions
are unique and depend on a harmonic function with a single center.
Hence the horizon geometry and the values of the relevant fields are
fully determined in terms of the charges. The hypermultiplet scalar
fields are covariantly constant but otherwise arbitrary. However, the
horizon and the entropy do not depend on these fields, so that they
can be ignored. In the absence of charges one is left with flat
Minkowski space-time with arbitrary constant moduli and
$T_{ab}{}^{ij}=0$.

As it turns out the attractor equations have a universal form. Before
commenting further on their derivation, let us present the equations,
which are manifestly covariant under electric/magnetic duality,
\begin{equation}
  \label{eq:BPS-attractor}
  \mathcal{P}^I=0\,,\qquad \mathcal{Q}_I= 0\,, \qquad \Upsilon=-64\,, 
\end{equation}
where 
\begin{eqnarray}
  \label{eq:P-Q-def}
  \begin{split}
  \mathcal{P}^I &\equiv p^I + \mathrm{i}(Y^I-\bar Y^I)  \,,
  \\ 
  \mathcal{Q}_I &\equiv q_I + \mathrm{i}(F_I-\bar F_I)  \,.
  \end{split}
\end{eqnarray}
Here the $Y^I$ and $\Upsilon$ are related to the $X^I$ and $A$,
respectively, by a uniform rescaling and $F_I$ and $F_\Upsilon$ will
denote the derivatives of $F(Y,\Upsilon)$ with respect to $Y^I$ and
$\Upsilon$.  To explain the details of the rescaling, we introduce the
complex quantity $Z$, sometimes refered to as the 'holomorphic BPS
mass', which equals the central charge associated with the vector
supermultiplets. In terms of the original variables $X^I$ it is
defined as
\begin{equation}
  \label{eq:Z}
    Z = \exp[\mathcal{K}/2] \, (p^I F_I(X,A) - q_I X^I) \,,
\end{equation}
where 
\begin{equation}
  \label{eq:cal-K}
    \mathrm{e}^{-\mathcal{K}} =  \mathrm{i} \,(\bar X^IF_I(X,A) - \bar
    F_I(\bar X,\bar A) X^I)\,. 
\end{equation}

At the horizon the variables $Y^I$ and $\Upsilon$ are defined by 
\begin{equation}
  \label{eq:Y-X-horizon}
  \begin{split}
  Y^I &= \exp[\mathcal{K}/2] \,\bar Z\,X^I \,,\\
  \Upsilon &= \exp[\mathcal{K}] \,\bar Z^2 \, A \,.
  \end{split}
\end{equation}
Note that $Y^I$ and $\Upsilon$ are invariant under arbitrary complex
rescalings of the underlying variables $X^I$ and $A$. The reader may
easily verify that for fields satisfying the attractor equations
\eqref{eq:BPS-attractor}, one establishes that
\begin{equation}
  \label{eq:Z2}
  \vert Z\vert^2 \equiv  p^I F_I - q_I Y^I \,,
\end{equation}
which is obviously real and positive, is equal to $\mathrm{i} (\bar
Y^IF_I -Y^I\bar F_I)$.

Finally we wish to draw attention to just one aspect of the derivation
of the attractor equations \eqref{eq:BPS-attractor}. Consider the
spinor fields belonging to the vector supermultiplets, and concentrate
on their supersymmetry variation in terms of a two-rank tensor,
\begin{equation}
  \label{eq:delta-Omega}
  \delta\Omega_i{}^I = \tfrac1{2} \varepsilon_{ij} \gamma^{ab}
  \epsilon^j 
  \Big(F_{ab}^-{}^I - \tfrac1{4} \varepsilon_{kl} T_{ab}{}^{kl} \,\bar
  X^I \Big) \,. 
\end{equation}
This particular linear combination of the field strength
$F_{ab}^-{}^I$ and the field $\bar X^I$ arises because the symmetry
transformations are evaluated in a superconformal background. Full
supersymmetry therefore implies that the right-hand side of
\eqref{eq:delta-Omega} vanishes, so that
\begin{equation}
  \label{eq:FproptoX}
  F_{ab}^-{}^I = \tfrac1{4} \varepsilon_{kl} T_{ab}{}^{kl} \,\bar X^I
  \,. 
  \end{equation}
An extension of this argument gives a similar result for the conjugate
field strengths, 
\begin{equation}
  \label{eq:GproptoF}
  G_{abI}^- = \tfrac1{4} \varepsilon_{kl} T_{ab}{}^{kl} \,\bar
  F_I\,.  
\end{equation} 
Note that these two equations are consistent with respect to
electric/magnetic duality. Given the fact that the field strengths
$F_{ab}{}^I$ and $G_{abI}$ satisfy the Maxwell equations and therefore
become proportional to the magnetic and electric charges, $p^I$ and
$q_I$, it is not surprising that one finds the attractor equations
\eqref{eq:BPS-attractor}. For further details of the analysis we refer
to \cite{LopesCardoso:2000qm}. 

\subsection{The BPS entropy function}
\label{sec:bps-entropy-function}
The BPS attractor equations follow also from a variational principle
based on the entropy function
\cite{Behrndt:1996jn,LopesCardoso:2006bg},
\begin{equation}
  \label{eq:Sigma-simple}
  \begin{split}
  &\Sigma(Y,\bar Y,p,q) =  \mathcal{F}(Y,\bar Y,\Upsilon,\bar\Upsilon)\\
  &\quad\qquad 
  - q_I   (Y^I+\bar Y^I) + p^I (F_I+\bar F_I)  \;,
  \end{split}
\end{equation}
where $p^I$ and $q_I$ couple to the corresponding magneto- and
electrostatic potentials at the horizon
(c.f.~\cite{LopesCardoso:2000qm}) in a way that is consistent with
electric/magnetic duality. The quantity $\mathcal{F}(Y,\bar
Y,\Upsilon,\bar\Upsilon)$, which will be regarded as a `free energy'
in what follows, is defined by
\begin{equation}
  \label{eq:free-energy-phase}
  \begin{split}
  \mathcal{F}(Y,\bar Y,\Upsilon,\bar\Upsilon)&= - \mathrm{i}\left(
  {\bar Y}^I F_I - Y^I {\bar F}_I
  \right) \\
  &\quad
  - 2\mathrm{i} \left( \Upsilon F_\Upsilon - \bar \Upsilon
  \bar F_{\Upsilon}\right)\,,
  \end{split}
\end{equation}
where $F_\Upsilon= \partial F/\partial\Upsilon$. Just as the entropy
function discussed in section \ref{sec:entropy-functions}, the entropy
function \eqref{eq:Sigma-simple} transforms as a function under
electric/magnetic duality \cite{deWit:1996ix}. Varying this entropy
function with respect to the $Y^I$, while keeping the charges and
$\Upsilon$ fixed, yields the result,
\begin{equation}
  \label{eq:Sigma-variation-1}
 \delta \Sigma = \mathcal{P}^I
 \, \delta ( F_{I} + \bar F_I)  -\mathcal{Q}_I\,
 \delta (Y^I+ \bar Y^I)  \;.
\end{equation}
Here we made use of the homogeneity of the function $F(Y,\Upsilon)$.
Under the mild assumption that the matrix 
\begin{equation}
  \label{eq:def-N}
  N_{IJ} = \mathrm{i}(\bar F_{IJ}-F_{IJ}), 
\end{equation}
is non-degenerate, it thus follows that stationary points of $\Sigma$
satisfy the attractor equations. The macroscopic entropy is equal to
the entropy function taken at the attractor point. This implies that
the macroscopic entropy is the Legendre transform of the free energy
\eqref{eq:free-energy-phase}.  An explicit calculation yields the
entropy formula \cite{LopesCardoso:1998wt,CarDeWMoh,LopesCardoso:1999ur},
\begin{equation}
  \label{eq:W-entropy}
  \begin{split}
  {\cal S}_{\rm macro}(p,q) &= \pi \, \Sigma\Big\vert_{\rm
  attractor}\\
  &= \pi \Big[\vert Z\vert^2 - 256\,
  {\rm Im}\, F_\Upsilon \Big ]_{\Upsilon=-64} \,.
  \end{split}
\end{equation}
Here the first term represents a quarter of the horizon area (in
Planck units) so that the second term defines the deviation from the
Bekenstein-Hawking area law.  The entropy coincides precisely with the
Wald entropy \cite{Wald:1993nt} as given by the right-hand side of
\eqref{eq:wald}. In fact, the original derivation of
\eqref{eq:W-entropy} was not based on an entropy function and made
direct use of the expression \eqref{eq:wald}.

In the absence of $\Upsilon$-dependent terms, the homogeneity of the
function $F(Y)$ implies that the area scales quadratically with the
charges, as was discussed already at the beginning of section
\ref{sec:macro-micro}. However, in view of the fact that $\Upsilon$
takes a fixed value, the second term will be subleading in the limit
of large charges.\footnote{ 
  Here one usually assumes that $F(Y,\Upsilon)$ can be expanded in
  positive powers of $\Upsilon$. }  
Note, however, that also the area will contain subleading terms, as it
depends on $\Upsilon$.

The entropy equation \eqref{eq:W-entropy} has been confronted with the
result of microstate counting, for instance, in the situation
described in section \ref{sec:M-bh}. In that case the effective
supergravity action is known and based on the function
\begin{equation}
  \label{eq:CY+cc-sg}
  \begin{split}
  F(Y,\Upsilon)&= -\frac1{6} \, \frac{C_{ABC}\,Y^AY^BY^C}{Y^0} \\
  &\quad
  - \frac{c_{2A}} {24\cdot 64}\, \frac{ Y^A}{Y^0} \,\Upsilon \,.
  \end{split}
\end{equation}
Substituting this result into \eqref{eq:W-entropy} and imposing the
attractor equations \eqref{eq:BPS-attractor} with $p^0=0$, one indeed
derives the result \eqref{eq:S-CY-micro} for the macroscopic entropy
\cite{LopesCardoso:1998wt}. The entropy formula \eqref{eq:W-entropy}
has also been put to a test in other cases. Some of them will be
discussed in due course.

The relation between \eqref{eq:Sigma-simple} and the entropy function
introduced in section \ref{sec:entropy-functions} was discussed in
\cite{CdWM}, where it was established that both entropy functions lead
to identical results for BPS black holes. When the black holes are not
BPS (i.e. have no residual supersymmetry), then the entropy function
\eqref{eq:Sigma-simple} is simply not applicable. In this connection
the question arises whether other, independent, higher-derivative
interactions associated with matter multiplets will not contribute to
the entropy either. For instance, Lagrangians for tensor
supermultiplets that contain interactions of fourth order in
space-time derivatives, lead to terms quadratic in the Ricci scalar
that will in principle contribute to the Wald entropy
\cite{deWit:2006gn}. Indeed, for non-BPS black holes these terms yield
finite contributions to the entropy, but for BPS black holes these
corrections vanish. A comprehensive treatment of higher-derivative
interactions is yet to be given for $N=2$ supergravity, but it seems
that this result is generic. At any rate, this observation seems in
line with more recent findings \cite{Exirifard:2006qv,Sahoo:2006pm}
based on heterotic string $\alpha^\prime$-corrections encoded in a
higher-derivative effective action in higher dimensions. In four
dimensions this action leads to additional matter-coupled
higher-derivative interactions.  When these are taken into account,
the matching of the macroscopic entropy with the microscopic result is
established \cite{Sahoo:2006pm}.

Another modification concerns possible non-holomorphic corrections to
the function $F(X,A)$. This holomorphic function leads to a
supersymmetric action that corresponds to the so-called effective
Wilsonian action, based on integrating out the massive degrees of
freedom. The Wilsonian action describes the correct physics for
energies between appropriately chosen infrared and ultraviolet
cutoffs. However, this action does not reflect all the physical
symmetries. To preserve those symmetries non-holomorphic contributions
should be included associated with integrating out massless degrees of
freedom. In the special case of heterotic black holes in $N=4$
supersymmetric compactifications, the requirement of explicit
S-duality invariance of the entropy and the attractor equations allows
one to determine the contribution from these non-holomorphic
corrections, as was first demonstrated in \cite{LopesCardoso:1999ur}
for BPS black holes. In \cite{Cardoso:2004xf,LopesCardoso:2006bg} it
was established that non-holomorphic corrections to the BPS entropy
function \eqref{eq:Sigma-simple} can be encoded into a real function
$\Omega(Y,\bar Y,\Upsilon,\bar\Upsilon)$ which is homogeneous of
second degree. The modifications to the entropy function are then
effected by substituting $F(Y,\Upsilon)\to F(Y,\Upsilon) + 2\mathrm{i}
\Omega(Y,\bar Y,\Upsilon,\bar\Upsilon)$. There are good reasons to
expect that this same substitution should be applied to the more
general entropy function based on \eqref{eq:entropy-function}
\cite{CdWM}. Note that when $\Omega$ is harmonic, i.e., when it
satisfies $\partial^2\Omega/\partial Y^I \,\partial\bar Y^J=0$, it can
simply be absorbed into the original holomorphic funtion
$F(Y,\Upsilon)$.

Finally we point out the existence of a formulation in terms of real,
rather than the complex fields $(Y^I, \Upsilon)$, that we used before.
This formulation is manifestly covariant with respect to
electric/magnetic duality. We first decompose $Y^I$ and $F_I$ into
their real and imaginary parts,
\begin{equation}
  \label{eq:u-y}
  Y^I = x^I + \mathrm{i} u^I \;,\qquad F_I = y_I + \mathrm{i} v_I \;,
\end{equation}
where $F_I = F_I(Y,\Upsilon)$.  The real parametrization is obtained
by taking $(x^I,y_I, \Upsilon, {\bar \Upsilon})$ instead of $(Y^I,
{\bar Y}^I, \Upsilon, {\bar \Upsilon})$ as the independent variables.
This reparametrization is only well defined provided
$\det(N_{IJ})\not=0$.  Subsequently one defines the Hesse potential,
the real analogue of the K\"ahler potential, which equals twice the
Legendre transform of the imaginary part of the prepotential with
respect to $u^I=\mbox{Im}\,Y^I$,
\begin{equation}
  \label{eq:GenHesseP}
  \mathcal{H}(x,y,\Upsilon, {\bar \Upsilon}) = 2 \; {\rm Im}\,F(x+\mathrm{i}
  u,\Upsilon,\bar\Upsilon)  - 2 \,  y_I \,u^I \;,
\end{equation}
Owing to the homogeneity of the function $F(Y,\Upsilon)$ one can show
that the free energy \eqref{eq:free-energy-phase} equals twice the Hesse 
potential. The entropy function (\ref{eq:Sigma-simple}) is now
replaced by
\begin{equation}
  \label{eq:RealSigma}
  \Sigma(x,y,p,q) = 2\,\mathcal{H}(x,y,\Upsilon,\bar\Upsilon) - 2 \,q_I
  x^I + 2\,p^I y_I \;,
\end{equation}
and it is straightforward to show that the extremization equations are
just the attractor equations \eqref{eq:BPS-attractor}, expressed in
terms of the new variables $(x^I,y_I)$. The value of $\Sigma(x,y,p,q)$
at the attrator point coincides again with the macroscopic entropy.

\subsection{Partial Legendre transforms}
\label{sec:partial-legendre}
It is, of course, possible to define the macroscopic entropy as a
Legendre transform with respect to only a subset of the fields, by
substituting part of the attractor equations such that the variational
principle remains valid.  These partial Legendre transforms constitute
a hierarchy of Legendre transforms for the black hole entropy. Here we
discuss two relevant examples, namely, the one proposed in
\cite{Ooguri:2004zv} where all the magnetic attractor equations are
imposed, and the dilatonic one for heterotic black holes, where only
two real potentials are left which together constitute the complex
dilaton field \cite{LopesCardoso:1999ur}. A possible disadvantage of
considering partial Legendre transforms is that certain invariances
may no longer be manifest. As it turns out, the dilatonic formulation
does not suffer from this shortcoming. Apart from that, there is no
reason to prefer one version over the other. This will change in
section~\ref{sec:partition-functions} when we discuss corresponding
partition functions and inverse Laplace transforms for the microscopic
degeneracies in semiclassical approximation.

Let us first impose the magnetic attractor equations so that only the
real parts of the $Y^I$ will remain as independent variables. Hence
one makes the substitution,
\begin{equation}
  \label{eq:yphi}
Y^I = \tfrac1{2}(\phi^I + \mathrm{i} p^I) \;.
\end{equation}
The entropy function (\ref{eq:Sigma-simple}) then takes the form,
\begin{equation}
  \label{eq:osv1}
  \Sigma(\phi,p,q) =  \mathcal{F}_{\rm
  E}(p,\phi,\Upsilon,\bar\Upsilon) - q_I   \,\phi^I  \,,
\end{equation}
where the corresponding free energy $\mathcal{F}_{\rm E}(p,\phi)$ equals
\begin{equation}
  \label{eq:osv2-nonholo}
  \begin{split}
  &
  \mathcal{F}_{\rm E}(p,\phi) = \\
  &\quad
  4\,\Big[ {\rm Im}\,F(Y,\Upsilon) +
  \Omega(Y,\bar Y,\Upsilon,\bar\Upsilon)\Big]_{Y^I=(\phi^I+ \mathrm{i}
  p^I)/2}  \,.
  \end{split}
\end{equation}
To derive this result one makes use of the homogeneity of the functions
$F(Y,\Upsilon)$ and $\Omega(Y,\bar Y,\Upsilon,\bar\Upsilon)$.  The
latter function may contain possible non-holomorphic terms.  When
extremizing (\ref{eq:osv2-nonholo}) with respect to $\phi^I$ we obtain
the attractor equations $q_I= \partial \mathcal{F}_{\rm E}/\partial
\phi^I$. This shows that the macroscopic entropy is a Legendre
transform of $\mathcal{F}_{\rm E}(p,\phi)$ subject to $\Upsilon=-64$,
as was first noted in \cite{Ooguri:2004zv} in the absence of $\Omega$.
In the latter case this Legendre transform led to the conjecture that
there is a relation with topological strings, in view of the fact that
$\exp[\mathcal{F}_{\rm E}]$ equals the modulus square of the
topological string partition function \cite{BCOV}. We return to this
in subsection \ref{sec:meas-mixed-part}. 

Along the same line one can now proceed and eliminate some of the
$\phi^I$ as well. A specific example of this is the dilatonic
formulation heterotic black holes, where we eliminate all the $\phi^I$
with the exception of two of them which parametrize the complex
dilaton field. This leads to an entropy function that depends only on
the charges and on the dilaton field
\cite{LopesCardoso:1999ur,Cardoso:2004xf}. Here it is convenient to
include all the $\Upsilon$-dependent terms into $\Omega$, which also
contains the non-holomorphic corrections. The heterotic classical
function $F(Y)$ is given by
\begin{equation}
  \label{eq:het-F0}
  F(Y) = - \frac{Y^1\,Y^a\eta_{ab} Y^b}{Y^0}\;,\qquad a = 2,\ldots,n,
\end{equation}
with real constants $\eta_{ab}$. In the application that we will be
considering the function $\Omega$ depends only linearly on $\Upsilon$
and $\bar\Upsilon$, as well as on the dilaton field $S=
-\mathrm{i}\,Y^1/Y^0$ and its complex conjugate $\bar S$. The result
for the BPS entropy function then takes the form,
\begin{eqnarray}
  \label{eq:nonholoSigma}
  \begin{split}
    \Sigma(S,\bar S,p,q) & =
   - \frac{q^2 - \mathrm{i} p\cdot q \, (S - {\bar S}) + p^2\,|S|^2}
    {S + {\bar S}}\\
    &\quad 
    + 4\, \Omega(S,\bar S,\Upsilon,\bar\Upsilon)\;,
    \end{split}
\end{eqnarray}
where $q^2$, $p^2$ and $p\cdot q$ are T-duality invariant bilinears of
the various charges, defined by
\begin{equation}
   \label{eq:chargeinvariants}
   \begin{split}
     q^2 &=  2 q_0 p^1 - \tfrac12  q_a \eta^{ab} q_b  \;,\\
     p^2 &= - 2  p^0 q_1 - 2 p^a \eta_{ab} p^b \;, \\
     q\cdot p &= q_0 p^0 - q_1 p^1 +  q_a p^a \;.
     \end{split}
\end{equation}
Note that these bilinears are not positive definite.  Furthermore,
$\Omega$ captures the $\Upsilon$-dependent corrections to the
classical result \eqref{eq:het-F0}. Its form was derived for $N=4$
heterotic string compactifications by requiring S-duality of the
attractor equations and of the entropy
\cite{LopesCardoso:1999ur,Cardoso:2004xf},
\begin{equation}
   \label{eq:Omega}
   \begin{split}
    & 
    \Omega(Y,\bar Y,\Upsilon,\bar \Upsilon) = \\
    &\quad
    \frac{1}{256\pi}\left[\Upsilon 
   \log \eta^{24}(S) + \bar\Upsilon \log \eta^{24}(\bar S)\right.\\
   &\quad\qquad\quad \left.
   +\tfrac1{2}(\Upsilon+ \bar \Upsilon) \log (S+\bar S)^{12} \right]
    \,,
    \end{split}
\end{equation}
where $\eta(S)$ denotes the Dedekind function.  Note the presence of
the last term which is non-holomorphic. This term is in accord with
the result for the effective action obtained from five-brane
instantons \cite{Harvey:1996ir}. The attractor equations associated
with the dilaton take the form $\partial_S\Sigma(S,\bar S,p,q)=0$.

It is interesting to consider the consequences of
\eqref{eq:nonholoSigma} in the classical case ($\Omega=0$). Then the
attractor equations yield the following values for the real part of
the dilaton field and the macroscopic entropy,
\begin{equation}
  \label{eq:dilaton-eq}
  \begin{split}
  S+\bar S &= - \frac2{p^2}\sqrt{p^2q^2-(p\cdot q)^2} \,,\\
  \mathcal{S}_{\mathrm{macro}} &= \pi \sqrt{p^2q^2-(p\cdot q)^2} \,.  
  \end{split}
\end{equation}
Obviously there are two types of black holes, depending on whether
$p^2q^2-(p\cdot q)^2$ is positive or zero. In the context of $N=4$
heterotic string compactifications these correspond to $1/4$- and
$1/2$-BPS black holes, respectively. The $1/4$-BPS states are dyonic,
so that they necessarily carry both electric and magnetic charges. The
$1/2$-BPS states can be purely electric. We derived these results from
the BPS entropy function, but they have also been obtained directly
from the full supergravity solutions
\cite{Cvetic:1995uj,Cvetic:1995,Bergshoeff:1996gg}.

Clearly the $1/4$-BPS black holes are {\it large} black holes as their
area (entropy) is nonzero and scales quadratically with the charges.
Note that, depending on the choice of the charges, the complex dilaton
field can remain {\it finite} in the limit of large charges. This is
relevant when studying the asymptotic growth of the dyonic degeneracy
of $1/4$-BPS dyons in heterotic string theory compactified on a
six-torus and in the related class of heterotic CHL models
\cite{Chaudhuri:1995fk}.  These degeneracies are encoded in
automorphic forms $\Phi_k (\rho, \sigma,\upsilon)$ of weight $k$ under
$\mathrm{Sp}(2,\mathbb{Z})$, or an appropriate subgroup thereof
\cite{Dijkgraaf:1996it,Jatkar:2005bh}. The torus compactification
corresponds to $k=10$ and the CHL models to $k=1,2,4,6$. The three
modular parameters, $\rho,\sigma,\upsilon$, parametrize the period
matrix of an auxiliary genus-two Riemann surface, which takes the form
of a complex, symmetric, two-by-two matrix. The microscopic degeneracy
of $1/4$-BPS dyons is expressed as an integral over an appropriate
3-cycle,
\begin{equation}
   \label{eq:dphik}
 d_k(p,q) = \oint \mathrm{d} \rho\,\mathrm{d}\sigma\,\mathrm{d}\upsilon \; 
 \frac{{\rm e}^{\mathrm{i} \pi [
     \rho\, p^2 + \sigma\, q^2 + (2 \upsilon -1)\, p \cdot q]}}
 {\Phi_k(\rho,\sigma,\upsilon)} \;.
\end{equation}
Since $\Phi_{k}$ has zeros in the interior of the Siegel half-space in
addition to the zeros at the cusps, the value of the integral
(\ref{eq:dphik}) depends sensitively on the choice of the integration
3-cycles. The charges are in general integer, with the exception of
$q_1$ which equals a multiple of $N$, and $p^1$ which is fractional
and quantized in units of $1/N$. Here $N$ and $k$ are related by
$(k+2)(N+1)=24$.  The inverse of the modular form $\Phi_k$ takes the
form of a Fourier sum with integer powers of $\exp[2\pi\mathrm{i}
\rho]$ and $\exp[2\pi\mathrm{i} \upsilon]$ and fractional powers of
$\exp[2\pi\mathrm{i}\sigma]$ which are multiples of $1/N$.  The
3-cycle is then defined by choosing integration contours where the
real parts of $\rho$ and $\upsilon$ take values in the interval
$(0,1)$ and the real part of $\sigma$ takes values in the interval
$(0,N)$.  The formula (\ref{eq:dphik}) is manifestly invariant under
T-duality (the integrand depends on the three T-duality invariant
bilinears (\ref{eq:chargeinvariants})), as well as under S-duality,
which is a subgroup of the full modular group.

The integral \eqref{eq:dphik} can be evaluated in saddle-point
approximation which yields the leading and subleading contribtions to
$d_k(p,q)$ \cite{Cardoso:2004xf,Jatkar:2005bh}. As it turns out these
contributions are precisely encoded in \eqref{eq:nonholoSigma} and
\eqref{eq:Omega}, including the Dedekind eta-functions and the
non-holomorphic terms. The presence of the non-holomorphic terms is
not surprising in view of the S-duality invariance. Note that the
expression \eqref{eq:Omega} refers to $k=10$ and that there exist
similar formulae for $k=1,2,4,6$.

The $1/2$-BPS black holes are {\it small} black holes as their area
scales linearly with the charges. According to \eqref{eq:dilaton-eq}
their entropy (and horizon area) vanishes while the dilaton field
diverges at the horizon, because we have $p^2=p\cdot q=0$. To describe
the situation more accurately one retains the leading term of
\eqref{eq:Omega}.  In that case one obtains the following result
(we restrict ourselves to $k=10$) \cite{LopesCardoso:1999ur},
\begin{equation}
  \label{eq:entcor-electric}
  \begin{split}
  S+\bar S &\approx  \sqrt{\vert q^2\vert/2}\,,\\
   \mathcal{S}_{\text{macro}} &\approx 4\,\pi\, \sqrt{\vert
   q^2\vert/2} -6\, \log{\vert q^2\vert}  \,,  
  \end{split}
\end{equation}
where the logarithmic term is due to the non-holomorphic contribution.
Because the dilaton is large in this case, all the exponentials in the
Dedekind eta-function are suppressed and we are at weak string
coupling $g_{\mathrm{s}} \propto (S+\bar S)^{-1/2}$.

We already stressed that small black holes have a size of the order of
the string scale and, indeed, these states are precisely generated by
perturbative heterotic string states arising in $N=4$ supersymmetric
compactifications to four space-time dimensions. In the supersymmetric
right-moving sector these states carry only momentum and winding and contain
no oscillations, whereas in the left-moving sector oscillations are
allowed that satisfy the string matching condition.  The oscillator
number is then linearly related to $q^2$. These perturbative states
received quite some attention in the past \cite{Dabholkar:1989jt}.
Because the higher-mass string states are expected to be within their
Schwarzschild radius, it was conjectured that they should have an
interpretation as black holes.\footnote{ 
  The idea that elementary particles, or string states, are behaving
  like black holes, has been around for quite some time.
} 
Their calculable level density, proportional to the exponent of $4\pi
\sqrt{\vert q^2\vert/2 }$, implies a nonzero microscopic entropy for
these black holes \cite{Russo:1994}. 

This result was confronted with explicit black hole solutions
\cite{Sen:1994,Sen:1995in,Peet:1995} based on standard supergravity
Lagrangians that are at most quadratic in derivatives, which have a
vanishing horizon area. Based on the area law one thus obtains a
vanishing macroscopic entropy. The fact that (\ref{eq:W-entropy}),
which takes into account higher-derivative interactions, can nicely
account for the discrepancies encountered in the classical description
of the 1/2-BPS black holes, was first emphasized in
\cite{Dabholkar:2004yr,Dabholkar:2004dq}. Note also that, since the
electric states correspond to perturbative heterotic string states,
their degeneracy is known from string theory and given by
\begin{equation}
  \label{eq:het-string}
  \begin{split}
    d(q) &= \oint {\mathrm d}\sigma\, \frac{{\mathrm
  e}^{\mathrm{i}\pi\sigma q^2}}{\eta^{24}(\sigma)} \\
  &\approx 
  \exp\left(4\pi\,\sqrt{|q^2|/2} - \tfrac{27}{4}
      \log \vert q^2\vert\right)  \,,
    \end{split}
\end{equation}
where the integration contour encircles the point $\exp(2\pi
\mathrm{i} \sigma)=0$. This large-$\vert q^2\vert$ approximation is
based on a standard saddle-point approximation. Obviously the leading
term of (\ref{eq:het-string}) is in agreement with
(\ref{eq:entcor-electric}, but beyond that there is a disagreement as
the logarithmic corrections carry different coefficients.  This
discrepancy may be regarded as a first indication that small black
holes are not well understood (for a disucssion, see, for instance,
\cite{LopesCardoso:2006bg}).  Therefore we will mainly concentrate on
large black holes in the next chapter.

\section{PARTITION FUNCTIONS AND INVERSE LAPLACE TRANSFORMS}
\label{sec:partition-functions}
To again make the connection with microstate degeneracies, we
conjecture, in the spirit of \cite{Ooguri:2004zv}, that the Legendre
transforms of the entropy are indicative of a thermodynamic origin of
the various entropy functions. It is then natural to assume that the
corresponding free energies are related to black hole partition
functions corresponding to suitable ensembles of black hole
microstates.  Following \cite{LopesCardoso:2006bg}, we define
\begin{equation}
  \label{eq:partition}
  Z(\phi,\chi) = \sum_{\{p,q\}} \;   d(p,q)
  \, \mathrm{e}^{\pi [q_I \phi^I - p^I \chi_I]} \,,
\end{equation}
where $d(p,q)$ denotes the microscopic degeneracies of the black hole
microstates with black hole charges $p^I$ and $q_I$. This is the
partition sum over a canonical ensemble, which is invariant under the
various duality symmetries, provided that the electro- and
magnetostatic potentials $(\phi^I,\chi_I)$ transform as a symplectic
vector. Identifying a free energy with the logarithm of
$Z(\phi,\chi)$, it is clear that it should, perhaps in an appropriate
limit, be related to the macroscopic free energy introduced earlier.
On the other hand, viewing $Z(\phi,\chi)$ as an analytic function in
$\phi^I$ and $\chi_I$, the degeneracies $d(p,q)$ can be retrieved by
an inverse Laplace transform,
\begin{equation}
  \label{eq:inverse-LT}
  d(p,q) \propto \int \;{\mathrm{d}\chi_I\,\mathrm{d}\phi^I}\;
  Z(\phi,\chi) \; \mathrm{e}^{\pi [-q_I \phi^I + p^I \chi_I]} \,,
\end{equation}
where the integration contours run, for instance, over the intervals
$(\phi-\mathrm{i}, \phi +\mathrm{i})$ and $(\chi-\mathrm{i}, \chi
+\mathrm{i})$ (we are assuming an integer-valued charge lattice).
Obviously, this makes sense as $Z(\phi,\chi)$ is formally periodic
under shifts of $\phi$ and $\chi$ by multiples of $2\mathrm{i}$.

These arguments suggest to identify $Z(\phi,\chi)$ with the Hesse
potential \eqref{eq:GenHesseP},
\begin{equation}
  \label{eq:partition-hesse}
  \begin{split}
   & 
   \sum_{\{p,q\}} \;   d(p,q)
  \,\mathrm{e}^{\pi[q_I \phi^I - p^I \chi_I]} \sim \\
  &\quad 
  \sum_{\rm shifts} \;
  \mathrm{e}^{2\pi\,\mathcal{H}(\phi/2,\chi/2,\Upsilon,\bar\Upsilon)}
  \,,
  \end{split} 
\end{equation}
where $\Upsilon$ is equal to its attractor value and where we
suppressed possible non-holomorphic contributions for simplicity.
However, the Hesse potential is a macroscopic quantity which does not
in general exhibit the periodicity that is characteristic for the
partition function.  Therefore, the right-hand side of
(\ref{eq:partition-hesse}) requires an explicit periodicity sum over
discrete imaginary shifts of the $\phi$ and $\chi$.\footnote{
  In case that the Hesse potential exhibits a periodicity with a
  different periodicity interval, then the sum over the imaginary
  shifts will have to be modded out appropriately such as to avoid
  overcounting. } 
When substituting $2\pi\mathcal{H}$ into the inverse Laplace
transform, we expect that the periodicity sum can be incorporated into 
the integration contour.

It is in general difficult to find an explicit representation
for the Hesse potential, as the relation (\ref{eq:u-y}) between the
complex variables $Y^I$ and the real variables $x^I$ and $y_I$ is
complicated. Therefore we rewrite \eqref{eq:partition-hesse} in terms
of the original variables $Y^I$ and $\bar Y^I$, where explicit results
are known,
\begin{equation}
  \label{eq:partition1}
  \begin{split}
   &
   \sum_{\{p,q\}} \;   d(p,q)
   \, \mathrm{e}^{\pi [q_I (Y+\bar Y)^I - p^I (\hat F +\hat{\bar
       F})_I]} \sim \\
   & \quad
   \sum_{\rm shifts} \;
  \mathrm{e}^{\pi \,\mathcal{F} (Y,\bar Y, \Upsilon,\bar \Upsilon) }
   \,. 
   \end{split}
\end{equation}
Here $\mathcal{F}$ equals the free energy \eqref{eq:free-energy-phase}
suitably modified with possible non-holomorphic corrections. The
latter requires that $F_I$ is changed into $\hat F_I= F_I +2
\mathrm{i} \Omega_I$, as was demonstrated in
\cite{LopesCardoso:2006bg}. It is important to note that both sides of
\eqref{eq:partition1} (as well as of \eqref{eq:partition-hesse}) are
manifestly consistent with duality.

Again, it is possible to formally invert (\ref{eq:partition1}) by
means of an inverse Laplace transform,
\begin{equation}
  \label{eq:laplace1}
  \begin{split}
  d(p,q) &\propto \int \; \mathrm{d} (Y+\bar Y)^I\; \mathrm{d} (\hat
  F+\hat{\bar F})_I 
  \;\mathrm{e}^{\pi\,\Sigma(Y,\bar Y,p,q)} \\
  &\propto \int \; \mathrm{d} Y\, \mathrm{d}\bar Y\;\Delta^-(Y,\bar Y)\;
  \mathrm{e}^{\pi\,\Sigma(Y,\bar Y,p,q)}\;,
  \end{split} 
\end{equation}
where $\Delta^-(Y,\bar Y)$ is an integration measure whose form
depends on $\hat F_I+ {\hat {\bar F}}_I$. The expression for
$\Delta^-$, as well as for a related determinant $\Delta^+$, reads as
follows,
\begin{equation}
  \label{eq:measure}
  \begin{split}
    &
    \Delta^\pm(Y,\bar Y) = \\
    &\quad
    \left\vert\det\Big[{\rm Im}\, F_{KL} + 2
      \,{\rm Re}(\Omega_{KL} \pm \Omega_{K{\bar L}}) \Big]\right\vert
    \,.  
    \end{split}
\end{equation}
As before, $F_{IJ}$ and $F_I$ refer to $Y$-derivatives of the
holomorphic function $F(Y,\Upsilon)$ whereas $\Omega_{IJ}$ and
$\Omega_{I\bar J}$ denote the holomorphic and mixed
holomorphic-antiholomorphic second derivatives of $\Omega$,
respectively. In the absence of non-holomorphic corrections
$\Delta^+=\Delta^-$.

A priori it is not clear whether the integral (\ref{eq:laplace1}) is
well-defined and we refer to \cite{LopesCardoso:2006bg} for a
discussion. Note that the periodicity sum in \eqref{eq:laplace1} is
defined in terms of the variables $\phi$ and $\chi$, which should have
some bearing on the integration contour in \eqref{eq:laplace1}.
Leaving aside these subtle points one may consider a saddle-point
approximation of the intergral representation (\ref{eq:laplace1}). In
view of the previous results it is clear that the saddle point
coincides with the attractor point. Subsequently one evaluates the
semiclassical Gaussian integral that emerges when expanding the
exponent in the integrand to second order in $\delta Y^I$ and $\delta
\bar Y^I$ about the attractor point.  When $Y^I-\bar
Y^I=\mathrm{i}p^I$, the resulting determinant factorizes into the
square roots of two subdeterminants, $\sqrt{\Delta^+}$ and
$\sqrt{\Delta^-}$. Here the plus (minus) sign refers to the
contribution of integrating over the real (imaginary) part of $\delta
Y^I$.  Consequently, the result of a saddle-point approximation
applied to (\ref{eq:laplace1}) yields,
\begin{equation}
  \label{eq:complex-saddle}
    d(p,q) = \sqrt{
  \left\vert \,\frac{\Delta^-(Y,\bar Y)}{\Delta^+(Y,\bar Y)}\right\vert }_{\rm
   attractor} \; \mathrm{e}^{{\cal S}_{\rm macro}(p,q)} \;.
\end{equation}
In the absence of non-holomorphic corrections the ratio of the two
determinants is equal to unity and one recovers precisely the
macroscopic entropy. In the presence of non-holomorphic terms, the
deviations from unity are usually suppressed in the limit of large
charges, and one recovers the leading and subleading corrections to
the entropy \cite{LopesCardoso:2006bg}. Of course, this is only the
case when the saddle-point approximation is appropriate. 

Alternatively one may choose to integrate only over the imaginary
values of the fluctuations $\delta Y^I$ in saddle-point approximation.
The saddle point then occurs in the subspace defined by the magnetic
attractor equations, so that one obtains a modified version of the OSV
integral \cite{Ooguri:2004zv},
\begin{equation}
  \label{eq:laplace-osv}
  d(p,q) \propto \int \; \mathrm{d} \phi \; \sqrt{\Delta^-(p,\phi)} \;
\mathrm{e}^{\pi[\mathcal{F}_{\rm E}(p,\phi)- q_I\phi^I]} \;,
\end{equation}
where $\mathcal{F}_{\rm E}(p,\phi)$ was defined in
(\ref{eq:osv2-nonholo}) and $\Delta^-(p,\phi)$ is defined in
(\ref{eq:measure}) with the $Y^I$ given by (\ref{eq:yphi}). Hence this
integral contains a non-trivial integration measure factor
$\sqrt{\Delta^-}$ in order to remain consistent with electric/magnetic
duality. Without the integral measure this is the integral conjectured
in \cite{Ooguri:2004zv}. In view of the original setting in terms of
the Hesse potential, we expect that the integration contours in
\eqref{eq:laplace-osv} should be taken along the imaginary axes. 

Inverting \eqref{eq:laplace-osv} to a partition sum over a mixed
ensemble, one finds,
\begin{equation}
  \label{eq:partition1-osv} 
  \begin{split}
    Z(p,\phi) &=
    \sum_{\{q\}} \; d(p,q) \, \mathrm{e}^{\pi \, q_I \phi^I } \\
    &\sim
  \sum_{\rm shifts} \;
  \sqrt{\Delta^-(p,\phi)}\;
    \mathrm{e}^{\pi\,\mathcal{F}_{\rm E}(p,\phi)}  \;.
   \end{split}
\end{equation}
It should be noted that this expression and the preceding one is less
general than (\ref{eq:laplace1}) because it involves a saddle-point
approximation. Moreover the function $\mathcal{F}_{\mathrm{E}}$ is not
duality invariant and the invariance is only recaptured when
completing the saddle-point approximation with respect to the fields
$\phi^I$. Therefore an evaluation of (\ref{eq:laplace-osv}) beyond the
saddle-point approximation will most likely give rise to a violation
of (some of) the duality symmetries again.

\subsection{The integration measure and the mixed partition function} 
\label{sec:meas-mixed-part}
As mentioned earlier, the partition function $Z(p,\phi)$ was
originally conjectured to be equal to the modulus square of the
partition function of the topological string \cite{Ooguri:2004zv}.
Soon thereafter, however, it was realized that this relationship must
be more subtle.  The arguments in \cite{LopesCardoso:2006bg}, based on
electric/magnetic duality clearly indicate some of the missing
ingredients, resulting in the measure factor $\sqrt{\Delta^-}$ and in 
the presence of the non-holomorphic corrections. In fact, it was
already clear from an early analysis of small heterotic black holes
that T-duality was not conserved when straightforwardly applying these
ideas \cite{Dabholkar:2005dt}. Although the presence of the measure factor
corrects for the lack of duality invariance, the semiclassical results
for small black holes seem to remain inconsistent with the analysis
of microstate counting, a fact we already alluded to at the end of
subsection \ref{sec:partial-legendre}.
 
It is possible to test the result \eqref{eq:partition1-osv} in the
context of the $1/4$-BPS states of the heterotic $N=4$ supersymmetric
string compactifications, by making use of the degeneracy formula
\eqref{eq:dphik}. Such a calculation was first performed in
\cite{Shih:2005he} and it was subsequently generalized in
\cite{LopesCardoso:2006bg} for more general charge configurations and
for CHL models. Using the degeneracies \eqref{eq:dphik} one calculates
the mixed partition sum on the left-hand side of
\eqref{eq:partition1-osv}. As it turns out, the resulting expression
is indeed proportional to the square of the partition function of the
topological string,
\begin{equation}
  \label{eq:z-top} 
  \vert Z_{\mathrm{top}}(p,\phi)\vert^2 =
  \mathrm{e}^{-2\pi\mathrm{i}[F(Y,\Upsilon) - 
  \bar F(\bar Y,\bar\Upsilon]} \,,
\end{equation}
where $F(Y,\Upsilon)$ is now the {\it holomorphic} part of \eqref{eq:het-F0}
and \eqref{eq:Omega}, 
\begin{equation}
  \label{eq:holo-F}
   \begin{split}
   F(Y,\Upsilon) &=
   - \frac{Y^1\,Y^a\eta_{ab} Y^b}{Y^0}\\
   &\quad
   + \frac{\mathrm{i}\,\Upsilon}{128\pi} \log\eta^{24}(-\mathrm{i}
   Y^1/Y^0)\;.  
  \end{split}
\end{equation}
Here $\Upsilon=-64$ and the $Y^I$ are given by
\eqref{eq:yphi}.\footnote{
  For convenience, we only refer to the case $k=10$. } 
However, there is a non-trivial proportionality factor, which, up to
subleading contributions, we expect to coincide with
\begin{equation}
  \label{eq:proportionality}
  \sqrt{\Delta^-(p,\phi)} \;\exp[4\pi\,\Omega_{\mathrm{nonholo}}] \,. 
\end{equation}
The expression for $\Omega_{\mathrm{nonholo}}$ follows from the last
term in \eqref{eq:Omega}, and is thus equal to 
\begin{equation}
  \label{eq:proportionality-nonholo}
   \exp[4\pi\;\Omega_{\mathrm{nonholo}}] = (S+\bar S)^{-12}\,, 
\end{equation}
where 
\begin{equation}
  \label{eq:Re-S-value}
  S+\bar S= \frac{2(p^1\phi^0-p^0\phi^1)}{(\phi^0)^2 + (p^0)^2} \;.  
\end{equation}
The factor $(S+\bar S)^{-12}$ cancels against a similar factor in
$\sqrt{\Delta^-}$ and, up to subleading terms,
\eqref{eq:proportionality} becomes
\begin{equation}
  \label{eq:proportionality-2}
  \begin{split}
    &\sqrt{\Delta^-(p,\phi)}
    \;\exp[4\pi\,\Omega_{\mathrm{nonholo}}]\approx \\ 
    &\quad
    \frac{\mathrm{i}(\bar Y^I\hat F_I - Y^I\hat F_I)}
    {2\, \vert Y^0\vert^2}= \frac{\mathrm{e}^{-\mathcal{K}(Y,\bar
    Y,\Upsilon,\bar\Upsilon)}}
    {2\, \vert Y^0\vert^2} \;.
  \end{split}
\end{equation}
Indeed this result coincides with the result found in
\cite{Shih:2005he,LopesCardoso:2006bg}. The expression for
$\mathrm{e}^{-\mathcal{K}}$ was already defined in \eqref{eq:cal-K} up
to non-holomorphic corrections. The latter can be dropped as they are
subleading. Note that $\mathrm{e}^{-\mathcal{K}}$ has been rescaled
by replacing the $X^I$ and $A$ by $Y^I$ and $\Upsilon$, respectively.
This is to be expected in view of the fact that the right-hand side of
\eqref{eq:proportionality-2} must be invariant under such rescalings.

However, we should discuss a subtlety related to the fact that we
derived the expression for $\sqrt{\Delta^-}$ in the context of $N=2$
supergravity, whereas the evaluation based on \eqref{eq:dphik} is
based on $N=4$ supersymmetric compactifications.  This means that the
number of scalar moduli (related to the number of $N=2$ vector
multiplets) is not obviously the same. In the case of $N=4$ the rank
of the gauge group is equal to 28 (for simplicity we restrict
ourselves to the case $k=10$). Of the 28 abelian vector gauge fields,
6 will correspond to the graviphotons of pure $N=4$ supergravity, and
22 will each belong to an $N=4$ vector supermultiplet. In the
reduction to $N=2$ supergravity, one of the graviphotons will be
contained in an additional $N=2$ vector supermultiplet and another one
will play the role of the graviphoton of $N=2$ supergravity.  There
are thus 24 abelian vector gauge fields, of which 23 are associated
with $N=2$ vector multiplets and one is associated with the $N=2$
graviphoton. The remaining 4 graviphotons are associated with two
$N=2$ gravitino supermultipets, and these vector fields seem to have
no place in the $N=2$ description.  Therefore it is often assumed
that the effective rank of the gauge group in the $N=2$ description
should be taken equal to 24, rather than to 28.

Nevertheless, in the calculation of $\sqrt{\Delta^-}$ leading to
\eqref{eq:proportionality-2} we assumed that the $N=2$ description is
based on 28 vector fields, corresponding to 27 vector supermultiplets
and one graviphoton field.  Only in that case, the factor $(S+\bar
S)^{-12}$ cancels so that one obtains the proportionality constant
noted in \eqref{eq:proportionality-2} based on the $N=2$ expression
for $\sqrt{\Delta^-}$. This somewhat confusing issue seems entirely
due to the fact that the $N=2$ description of an $N=4$ theory is not
fully understood, as the corresponding description of $N=4$
supergravity is indeed based on 28 electrostatic potentials $\phi$
which are contained in 28, rather than 24, analogues of the $N=2$
quantities $Y^I$.

In a recent series of papers
\cite{Gaiotto:2006ns,Beasly:2006us,deBoer:2006vg} progress was made
towards a further understanding of the relation between the mixed
black hole partition function and the partition function of the
topological string. Unfortunately no evidence for the presence of the
integration measure factor in \eqref{eq:laplace-osv} and
\eqref{eq:partition1-osv} was presented. However, a more detailed
analysis for compact Calabi-Yau \cite{Denef:2007} models subsequently
revealed the presence of the measure factor. Based on an extensive
analysis of the factorization formula for BPS indices, it is shown
that the partition function does not completely factorize into a
holomophic and an anti-holomorphic sector and the measure that is
found agrees (for $p^0=0$) with \eqref{eq:proportionality-2}. The
power of $\vert Y^0\vert$ depends on whether one is discussing $N=2$
or $N=4$ black holes. These results seem to be in line with what was
discussed in this section, although there are many subtleties.
Obviously more work is needed to fully explore their consequences.

\vspace{2ex}
\noindent
The work described in these lectures is based on various
collaborations with Gabriel Lopes Cardoso, J\"urg K\"appeli, Swapna
Mahapatra, Thomas Mohaupt and Frank Saueressig. I thank Gabriel Lopes
Cardoso for discussion and valuable comments on the manuscript.\\
This work is partly supported by NWO grant 047017015, EU contracts
MRTN-CT-2004-005104 and MRTN-CT-2004-512194, and INTAS contract 03-51-6346.


\end{document}